\documentclass[twocolumn,aps,prb,showpacs,groupedaddress]{revtex4-1}
\usepackage{amssymb,amsbsy,graphicx,xcolor,subfigure,marvosym,color,bm}
\usepackage{amsmath}
\usepackage{graphicx}

\usepackage{pifont}

\usepackage{amsfonts}
\usepackage{subfigure}
\usepackage{setspace}
\usepackage{threeparttable}
\usepackage{enumerate}
\usepackage{xspace}
\usepackage{textcomp}
\usepackage{bm}
\usepackage{graphicx,color,epsfig}

\begin{document}

\title{Real space Berry curvature of itinerant electron systems with spin-orbit interaction}

\author{Shang-Shun Zhang$^{1}$, Hiroaki Ishizuka$^{2}$, Hao Zhang$^{1,3}$, G\'abor B. Hal\'asz$^{3}$,  and Cristian D. Batista$^{1,4}$}

\address{$^1$Department of Physics and Astronomy, University of Tennessee, Knoxville,
Tennessee 37996-1200, USA}
\address{$^2$Department of Applied Physics, The University of Tokyo, Hongo, Bunkyo, Tokyo, 113-8656, Japan}
\address{$^3$Materials Science and Technology Division, Oak Ridge National Laboratory, Oak Ridge, Tennessee 37831, USA}
\address{$^4$Quantum Condensed Matter Division and Shull-Wollan Center, Oak Ridge National Laboratory, Oak Ridge, Tennessee 37831, USA}

\begin{abstract}
By considering an extended double-exchange model with spin-orbit
coupling (SOC), we derive a general form of the Berry phase $\gamma$
that electrons pick up when moving around a closed loop. This form
generalizes the well-known result valid for SU(2) invariant systems,
$\gamma=\Omega/2$, where $\Omega$ is the solid angle subtended by
the local magnetic moments enclosed by the loop. The general form of
$\gamma$ demonstrates that collinear and coplanar magnetic textures
can also induce a Berry phase different from 0 or $\pi$, smoothly
connecting the result for SU(2) invariant systems with the
well-known result of Karplus and Luttinger for
collinear ferromagnets with finite SOC. By taking
the continuum limit of the theory, we also derive the corresponding
generalized form of the real space Berry curvature. The new
expression  is a generalization of the scalar spin chirality, which
is presented in an explicitly covariant form. We finally show how
these simple concepts can be used to understand the
origin of the spontaneous topological Hall effect that has been
recently reported in collinear and coplanar antiferromagnetic phases
of correlated materials.
\end{abstract}
\maketitle

\section{Introduction} \label{sec-int}

The phenomenon of colossal magnetoresistance (CMR) provides a clear
example of the dramatic effects of magnetism on electronic
transport.~\cite{Dagotto11,Tokura2000} The ability of changing the
longitudinal  electric resistivity by several orders of magnitude
with an external magnetic field generated a wide
interest,~\cite{Ramirez97,Rodriguez-Martinez96} not only due to its
multiple technological applications, including magnetic recording,
but also because of the rich fundamental physics arising from the
interplay between charge and spin degrees of freedom. Indeed,
colossal magnetoresistance is an example of the profound influence
of quantum mechanics on the macroscopic behavior of correlated
materials.

The phenomenon of CMR only refers to the diagonal components of the
conductivity tensor. It is natural to ask if a
similar  dramatic change of the {\it off-diagonal components of the
conductivity tensor} can also be achieved by exploiting the
interplay between localized magnetic moments and conduction
electrons.~\cite{Nagaosa06} Like in the case of CMR materials, the
first affirmative answer to this question  arose from the study of
ferromagnets, whose transverse resistivity contains a term that
remains non-zero even after switching off the applied magnetic
field.~\cite{Hall1881,Smith21,Hurd72} Given that the magnetization,
${\bm M}$, is apparently the only axial vector that characterizes
the system in absence of the external field, it is not surprising
that the observed ``anomalous Hall effect" (AHE) turned out to be
proportional to ${\bm M}$. However, subsequent experiments showed
that $\rho_{xy}$ can be a non-monotonic function of $M$. This key
observation led to the discovery of a much more interesting
phenomenon of quantum mechanical origin.~\cite{Taguchi01}

Karplus and Luttinger made a seminal contribution to the problem of
anomalous Hall effect by considering the band structure of
ferromagnets in the presence of spin-orbit
interaction.~\cite{Karplus54} They pointed out that  the anomalous
velocity arises from inter-band matrix elements of the current
operator. Smit provided an alternative  explanation of the
phenomenon by attributing the AHE to skew scattering with
impurities.~\cite{Smit58} Later, another extrinsic mechanism,
side jump, was proposed by Berger, which is related to the shift of
the electron during its collision with an
impurity.~\cite{Berger70} The skew scattering mechanism leads to a
contribution to $\rho_{xy}$ that is proportional to
the longitudinal resistivity $\rho$. In contrast, the intrinsic
and side-jump mechanisms give $\rho_{xy}
\propto \rho^2$. This behavior of $\rho_{xy}$ was
confirmed by a transport measurement in the spinel
CuCe$_2$Se$_{4-x}$Br$_{x}$.~\cite{Lee04} In general, the dominant
mechanism depends on the material consideration. Our current
understanding indicates that skew scattering is dominant in
relatively clean materials, while the intrinsic mechanism is
dominant in relatively dirty materials.~\cite{Miyasato07,Onoda08}

The vast implications of the explanation of the AHE offered by
Karplus and Luttinger  became clear after the derivation of the Hall
conductance by Thouless {\emph{et.~al.}~and their
analysis of the quantum Hall effect.~\cite{Thouless82} In
particular, it was understood that the AHE can be attributed to a
Berry phase~\cite{Berry84} associated with the Bloch wave functions
in solids.~\cite{Onoda02,Jungwirth02} This Berry phase can arise
through different mechanisms. Ye
{\emph{et.~al.}~\cite{Ye99} proposed a mechanism
based on the observation that a carrier moving in a {\it
non-coplanar} spin background acquires a real space Berry phase,
which affects the motion of electrons in the same way
as the Aharonov-Bohm~\cite{Aharonov59} phase
arising from a physical magnetic field. Refs.~\onlinecite{Matl98}
and \onlinecite{Chun00} also discussed the relevance of non-coplanar
spin configurations to the AHE  in the context of perovskite-type
manganites at high temperatures. A similar mechanism was proposed
for pyrochlore ferromagnets~\cite{Ohgushi00,Taguchi01} and for
noncoplanar
antiferromagnets.~\cite{Shindou01,Martin08,Kato10,Akagi10,Barros14,Batista16,Ozawa17}
However, Jungwirth et. al.~\cite{Jungwirth02} showed that
non-coplanar spin ordering is
not actually necessary to produce an AHE. They
related the  AHE of {\it collinear ferromagnets}
directly to a Berry phase in momentum space, which arises from the
way in which the  spin-orbit coupled Bloch wave functions depend on
the wave vector. In simple terms, the net spin magnetization of a
ferromagnetic system induces an orbital current (or orbital magnetic
moment)  via the spin-orbit interaction. The combination of a net
spin magnetization and SOC then  leads to an effective magnetic
field which couples to the orbital degrees of freedom of the (spin
polarized) conduction electrons.

In the case of collinear and coplanar magnetic orderings, the SOC
can only arise from the {\it relativistic spin-orbit
interaction}.~\cite{Chen14} In contrast, the beauty of non-coplanar
magnetic orderings is that  they produce an effective SOC in absence
of any relativistic contribution. This phenomenon arises from the
Berry  phase $\gamma$ acquired by the electronic wave function  when
the electron moves in a closed  loop, which cannot be distinguished
from the Aharonov-Bohm phase~\cite{Aharonov59} produced by a
magnetic flux equal to $\gamma \Phi_0/2 \pi$ ($\Phi_0$ is the flux
quantum). Given that spins of an elementary plaquette can subtend a
solid angle comparable to $2\pi$, the effective magnetic flux
produced by a non-coplanar spin ordering can be of the order of one
flux quantum per elementary plaquette. For real materials, the area
of an elementary plaquette can be as small as a few {\AA}$^2$,
implying that the effective magnetic field produced by a
non-coplanar spin ordering can reach values of order $10^5$T. This
is an enormous magnetic field  if we consider that the strongest
pulsed field that can be currently generated in the high magnetic
field facilities are slightly higher than 100T.

The effective gauge field induced by non-coplanar
spin-orderings~\cite{Wen89,Nagaosa90,Lee92} can be illustrated with
the so-called $s$-$d$ exchange model. This idea originates  from the
seminal papers by Zener,\cite{Zener51} and Anderson and
Hasegawa,~\cite{Anderson55} which discuss the
double exchange interaction  generated in the limit of large
exchange coupling between the local moments and the conduction
electrons. In this limit, the spin of the electron is forced to be
aligned with the underlying spin field. As a consequence, the
effective hopping matrix element between two sites $j$ and $k$ with
local classical magnetic moments ${\bm S}_j= S {\bm n}_j$ and ${\bm
S}_k= S {\bm n}_k$, becomes
\begin{equation}
{\tilde t}_{kj} = t_{kj} \langle {\bm n}_j | {\bm n}_k \rangle = t_{kj} \cos{(\chi_{kj}/2)} e^{i \Omega_{njk}/2},
\label{dem}
\end{equation}
where $| {\bm n}_k \rangle$ is the coherent spin state polarized
along the ${\bm n}_k$ direction, ${\bm S}_k \cdot {\bm n}_k | {\bm
n}_k \rangle = S | {\bm n}_k \rangle$, $\chi_{kj}$ is the angle
between the $j$ and $k$ local moments,
$\chi_{kj}=\arccos({{\bm n}_j \cdot {\bm n}_k})$,
and $\Omega_{njk}$ is the solid angle subtended by ${\bm n}_j $,
${\bm n}_k$, and a reference unit vector ${\bm n}$. When the
electron moves in a closed loop, such as a triangular plaquette, it
picks up a Berry phase equal to the sum of the phases $\Omega_{njk}$
of each hopping amplitude. This sum  is independent of the reference
vector ${\bm n}$ (gauge invariance) and equal to half of the solid
angle enclosed by the loop of spins on the unit sphere. In the
continuum limit (infinitesimal small loops), this solid angle is
proportional to the scalar spin chirality, ${\bm n}_j \cdot {\bm
n}_k \times {\bm n}_l$, which then acts as an effective magnetic
field on the orbital degree of freedom of the conduction
electrons.~\cite{Volovik03}

In general,  the spin-orbit interaction, which is always present in solids, modifies the above argument based on  the  scalar spin chirality. For example,  a theoretical study of Mo oxides has proposed that the spin-orbit interaction significantly modifies the Hall conductivity of the non-coplanar phase, producing a much larger Hall effect in multi-band systems.~\cite{Tomizawa09} However, to date, a systematic understanding of how the spin-orbit interaction modifies  the geometric picture associated with the scalar spin chirality mechanism is lacking. The purpose of this article is to derive the
effective gauge field that emerges from a given magnetic texture
{\it in the presence of finite SOC}. The SOC enters in the
single-electron tight-binding Hamiltonian as a fixed SU(2) gauge
field (defined on the lattice bonds),~\cite{Frohlich93,Tokatly08}
whose value is determined by the interplay between the relativistic
spin-orbit interaction and the crystal structure. We note that this
field can  fluctuate in theories where the ionic positions are
allowed to fluctuate (e.g., theories that include electron-phonon
coupling). In this work, we only include the electronic degrees of
freedom,  implying that the SU(2) gauge field remains frozen. The
inclusion of finite SOC coupling in the hopping term of the $s$-$d$
model leads to a more general form of the real space Berry curvature
(or effective magnetic field) in the double-exchange limit. In
particular, it will become evident that non-coplanarity of the
magnetic structure is no longer a requirement for producing a
non-trivial real space Berry curvature, i.e., collinear and coplanar
structures can also produce an effective magnetic field in
the presence of SOC. We will also see that the case
of collinear ferromagnets that was originally considered by Karplus
and Luttinger~\cite{Karplus54} is simply a limiting case of the
general formula that we will derive here. Moreover, by taking the
continuum limit of the model Hamiltonian under consideration, we
derive an explicitly gauge invariant form of the Berry
curvature that generalizes the notion of scalar spin chirality.
Finally, we consider a few simple examples to illustrate the
applicability of these simple concepts to models and materials that
exhibit topological Hall effect induced by coplanar and collinear
magnetic orderings.

The presentation of our results is organized as follows. In
Sec.~\ref{sec-gen} we formulate the general problem and we introduce
the model Hamiltonian that will be used in the rest of the
manuscript. Sec.~\ref{sec-geom} introduces a geometric approach for
the computation of the electronic Berry phase in the presence of
SOC. Mathematically oriented readers may find this approach more
appealing than the algebraic treatment that is introduced in
Sec.~\ref{sec-alg}. However, the algebraic approach is probably
more amenable for the general reader, who can skip
Sec.~\ref{sec-geom} in a first reading of the
manuscript. A similar consideration can be applied to the continuum
limit of the theory that is described in
Sec.~\ref{sec-cont}. While it is important to understand how the
notion of scalar spin chirality (i.e., the source
of the Berry curvature in the SU(2) invariant case)
must be generalized in the presence of SOC, this is not a
requirement for understanding the subsequent
section, which is devoted to applying the
generalized form of the real space Berry curvature to
simple lattice models. These simple models in
Sec.~\ref{sec-mom} capture the essence of the topological Hall
effect that has been recently observed in materials with coplanar
and collinear antiferromagnetic orderings, such as
Mn$_3$Sn~\cite{nakatsuji2015large} and
CoNb$_3$S$_6$.~\cite{Ghimire18}

\section{General formulation} \label{sec-gen}

We will consider itinerant electrons that interact
with localized magnetic moments via an exchange coupling $J$. For
simplicity, we will  assume that quantum fluctuations are small,
implying the localized spins can be approximated by classical
moments, ${\bm S}_j = S {\bm  n}_j$, where ${\bm  n}_j$ is a
normalized vector field,
\begin{equation}
{\bm  n}_j = \left( \sin \theta_j \cos \phi_j, \sin \theta_j \sin
\phi_j, \cos \theta_j \right), \label{eq-gen-s-1}
\end{equation}
representing  the direction of the classical moment ${\bm  S}_j$. In
this classical limit, the sign of the exchange interaction $J$ can
be changed by the unitary transformation: ${\bm  S}_j \to - {\bm
S}_j$. Consequently, without loss of generality, we will adopt a
ferromagnetic sign $J>0$. The simplest Hamiltonian that describes
this physics is:
\begin{eqnarray}
{\cal H} &=& {\cal H}_t + {\cal H}_J,
\nonumber \\
{\cal H}_{t} &=& \sum_{jk}  (t_{kj} {\bm {c}}_{k}^{\dagger} U_{kj} {\bm {c}}_{j}+ t^*_{kj}  {\bm {c}}_{j}^{\dagger} U^{\dagger}_{kj} {\bm {c}}_{k}  ),\label{kin}
\nonumber \\
{\cal H}_{J} &=& - \frac{J S}{2} \sum_j {\bm {c}}_{j}^{\dagger} {\boldsymbol \sigma} {\bm {c}}_{j}^{\;}\cdot {\bm  n}_j.
\label{Hamiltonian}
\end{eqnarray}
where ${\boldsymbol \sigma} = (\sigma_1, \sigma_2, \sigma_3)$ is a vector of the Pauli matrices. Here we are using the spinor notation:
\begin{equation}
{\bm {c}}_{k}=\left[{\begin{array}{c}
c_{k\uparrow}\\
c_{k\downarrow}
\end{array}}\right], \;\;\;{\bm {c}}_{k}^{\dagger}=\left[{\begin{array}{cc}
c_{k\uparrow}^{\dagger} & c_{k\downarrow}^{\dagger}\end{array}}\right].
\end{equation}
The unitary operator $U_{kj}$ is  an SU(2) rotation matrix taking the general form
\begin{equation}
U_{kj} = \exp \left[ -\frac{i \alpha_{kj}} {2} \left( {\bm  a}_{kj}
\cdot {\boldsymbol \sigma} \right) \right], \label{eq-gen-U}
\end{equation}
where $\alpha_{kj}$ is the rotation angle induced by the finite
spin-orbit interaction, ${\bm  a}_{kj}$ is the unit vector in the
direction of the rotation axis.
The complex hopping amplitudes $t_{kj}$ can be expressed as $t_{kj}
= |t_{kj}| e^{i \beta_{kj}} $. We note that ${\cal H}$ becomes SU(2)
invariant  in absence of SOC: $\alpha_{kj}=0  \, (\forall kj)$.

We will also assume that the exchange interaction $J$ is comparable
or larger than the bandwidth of the itinerant electrons. In this
so-called double-exchange limit,  the electronic spin orientation
must remain parallel to the underlying localized spin. The resulting
low-energy Hamiltonian is simply a spinless fermion tight-binding
model with effective complex hopping amplitudes,
\begin{equation}
{\tilde t}_{kj} = \tau_{kj}  e^{i \gamma_{kj}},
\end{equation}
determined by the configuration of the underlying magnetic moments:
\begin{equation}
\tau_{kj}  = t_{kj} \sqrt{\frac{1 + {\bm n}_k \cdot R_{kj} \cdot
{\bm n}_j}{2}}.
\end{equation}
The SO(3) rotation matrix
\begin{equation}
R_{kj} = \exp \big[ \alpha_{kj} \big( {\bm a}_{kj} \cdot {\bm L}
\big) \big], \label{eq-soc-R}
\end{equation}
corresponds to the SU(2) rotation defined by the matrix $U_{kj}$ in
Eq.~(\ref{eq-gen-U}), where ${\bm L} = (L^x, L^y,
L^z)$ is a vector of the standard SO(3) generators  $[L^a]_{bc} =-
\varepsilon^{abc}$. This is just a generalization of the
Anderson-Hasegawa or Double Exchange model to the case with finite
SOC.~\cite{Zener51,Anderson55}

We are interested in the net phase $\Phi_{jkl}$ that the electronic
wave function picks up as the electron moves around the triangle
$jkl$. This phase is equal to the sum of two contributions:
\begin{equation}
\Phi_{jkl} = \beta_{jkl} + \gamma_{jkl},
\end{equation}
where $\beta_{jkl} = \beta_{jl} + \beta_{lk} + \beta_{kj}$,
is the phase that arises from the complex nature of the hopping amplitudes $t_{kj}$, which must  be equal to
$0$ or $\pi$ because ${\cal H}_t$ is time reversal invariant. The Berry phase,
\begin{equation}
\gamma_{jkl} = \gamma_{jl} + \gamma_{lk} + \gamma_{kj},
\label{eq-gen-gamma-1}
\end{equation}
arises from the strong exchange interaction between the electronic
spin and the local moments, i.e., from the projection of the
electronic spin state of each site $j$ into the low-energy state:
\begin{equation}
| {\bm n}_j \rangle = \cos \frac{\theta_j} {2} \, | \uparrow \,
\rangle + \sin \frac{\theta_j} {2} \, e^{i \phi_j} | \downarrow \,
\rangle.
\label{eq-gen-s-2}
\end{equation}
While our requirement of adiabaticity is not necessary for
generating a real space Berry curvature, it greatly simplifies the
analysis. In particular, it helps to identify  the Berry curvature
with a fictitious  magnetic field that couples to the orbital motion
of the itinerant electrons. In the absence of SOC,
the flux of this effective magnetic field (in units of the flux
quantum) through a closed loop is equal to  half of the solid angle
subtended by the local moments when moving around that loop.
Correspondingly, in the long wavelength limit, the fictitious
magnetic field on a triangular plaquette $jkl$ is  proportional to
the scalar product of the three local moments:
${\bm n}_j \cdot {\bm n}_k \times {\bm n}_l$.

In the presence of finite SOC, the Berry connection
$\gamma_{kj}$ becomes
\begin{equation}
\gamma_{kj} = \mathrm{arg} \left[ \langle {\bm  n}_k | U_{kj} |
{\bm  n}_j \rangle \right], \label{eq-gen-gamma-2}
\end{equation}
where $U_{kj}$ is the unitary operator
in Eq.~(\ref{eq-gen-U}) describing
the spin rotation as the electron hops from $j$ to
$k$. The goal of the next sections is to understand how this {\it
extra spin rotation induced by finite SOC} modifies the effective
magnetic field generated by the underlying vector field ${\bm n}$.
Along this process, we will find an explicitly covariant expression
for this effective magnetic field and we will learn that such a
field can be non-zero for {\it collinear} and {\it coplanar}
magnetic orderings, in addition to the non-coplanar orderings that
are required for SU(2) invariant systems. We will  see that this
result unifies under a common frame the different magnetic
orderings, such as collinear ferromagnetism and non-coplanar
antiferromagnetism, that were previously identified as
distinct potential sources of topological Hall
effect.

\section{Geometric approach}
\label{sec-geom}

\subsection{Geodesic spin rotations} \label{sec-rot}

For any two spin states $| {\bm p} \rangle$ and $| {\bm q} \rangle$
corresponding to non-collinear unit vectors ${\bm p}$ and ${\bm q}$,
we can define a ``geodesic'' SU(2) rotation $\tilde{U}_{{\bm q},
{\bm p}}$ that rotates the spin direction from ${\bm p}$ to ${\bm
q}$ along a geodesic (i.e., a great circle) of the Bloch sphere. If
we choose the geodesic spin rotation to be along the shorter arc of
the great circle containing both ${\bm  p}$ and ${\bm q}$, it is
unique and is given by
\begin{equation}
\tilde{U}_{{\bm  q}, {\bm  p}} = \exp \left[ -\frac{i \chi_{{\bm  q},
{\bm  p}}} {2} \left( {\bm  u}_{{\bm  q}, {\bm  p}} \cdot {\boldsymbol \sigma}
\right) \right],
\label{eq-rot-U}
\end{equation}
where the rotation angle $\chi_{{\bm q}, {\bm p}} = \arccos ({\bm p}
\cdot {\bm q}) < \pi$ is the angle between the two unit vectors
${\bm p}$ and ${\bm q}$, and the unit vector specifying the rotation
axis is
\begin{equation}
{\bm u}_{{\bm q}, {\bm p}} = \frac{{\bm p} \times {\bm q}} {|{\bm p}
\times {\bm q}|} = \frac{{\bm p} \times {\bm q}} {\sin \chi_{{\bm
q}, {\bm p}}}. \label{eq-rot-u}
\end{equation}
It is instructive to consider the matrix element of this geodesic
spin rotation $\tilde{U}_{{\bm q}, {\bm p}}$ between the two spin
states $| {\bm p} \rangle$ and $| {\bm q} \rangle$. By construction,
$\tilde{U}_{{\bm q}, {\bm p}} | {\bm p} \rangle \propto | {\bm q}
\rangle$, and the geodesic matrix element $\langle {\bm q} |
\tilde{U}_{{\bm q}, {\bm p}} | {\bm p} \rangle$ is thus a complex
number of unit modulus. To determine its complex argument, we first
expand $\tilde{U}_{{\bm q}, {\bm p}}$ in the standard way:
\begin{eqnarray}
\langle {\bm q} | \tilde{U}_{{\bm q}, {\bm p}} | {\bm p} \rangle &=&
\langle {\bm q} | \left[ \cos \frac{\chi_{{\bm q}, {\bm p}}} {2} - i
\left( {\bm u}_{{\bm q}, {\bm p}} \cdot {\boldsymbol \sigma} \right)
\sin \frac{\chi_{{\bm q}, {\bm p}}} {2} \right] | {\bm p} \rangle
\nonumber \\
&=& \cos \frac{\chi_{{\bm q}, {\bm p}}} {2} \, \langle {\bm q} |
{\bm p} \rangle
\nonumber \\
&-& \frac{i}{2} \left[ \cos \frac{\chi_{{\bm q}, {\bm p}}} {2}
\right]^{-1} \langle {\bm q} | \left[ \left( {\bm p} \times {\bm q}
\right) \cdot {\boldsymbol \sigma} \right] | {\bm p} \rangle.
\label{eq-rot-qp-1}
\end{eqnarray}
With some straightforward algebra, it can then be shown that the
matrix element in the second term of Eq.~\eqref{eq-rot-qp-1} is proportional to the overlap in
the first term:
\begin{equation}
\langle {\bm q} | \left[ \left( {\bm p} \times {\bm q} \right) \cdot
{\boldsymbol \sigma} \right] | {\bm p} \rangle = 2i \sin^2
\frac{\chi_{{\bm q}, {\bm p}}} {2} \, \langle {\bm q} | {\bm p}
\rangle. \label{eq-rot-qp-2}
\end{equation}
Using this relation, the geodesic matrix element in
Eq.~(\ref{eq-rot-qp-1}) can be written as
\begin{equation}
\langle {\bm q} | \tilde{U}_{{\bm q}, {\bm p}} | {\bm p} \rangle =
\left[ \cos \frac{\chi_{{\bm q}, {\bm p}}} {2} \right]^{-1} \langle
{\bm q} | {\bm p} \rangle = \frac{\langle {\bm q} | {\bm p} \rangle}
{|\langle {\bm q} | {\bm p} \rangle|}. \label{eq-rot-qp}
\end{equation}
Since $|\langle {\bm q} | {\bm p} \rangle|$ is real and positive by
definition, the argument of the geodesic matrix element $\langle
{\bm q} | \tilde{U}_{{\bm q}, {\bm p}} | {\bm p} \rangle$ is
identical to the argument of the overlap $\langle {\bm q} | {\bm p}
\rangle$.

\vspace{1cm}

\subsection{SU(2) invariant case}
\label{sec-inv}

If the electron spin is not rotated as it hops between different
sites of the triangle, the Berry connection in
Eq.~(\ref{eq-gen-gamma-2}) takes the simplified form
\begin{equation}
\gamma_{kj} = \mathrm{arg} \left[ \langle {\bm n}_k | {\bm n}_j
\rangle \right]. \label{eq-inv-gamma-1}
\end{equation}
Employing Eq.~(\ref{eq-rot-qp}) to turn the overlap in
Eq.~(\ref{eq-inv-gamma-1}) into a geodesic matrix element, this
Berry connection can be written as
\begin{eqnarray}
\gamma_{kj} &=& \mathrm{arg} \left[ \langle {\bm n}_k |
\tilde{U}_{{\bm n}_k, {\bm n}_j} | {\bm n}_j \rangle \right] =
\mathrm{arg} \Big[ \langle {\bm n}_k | \left( \tilde{U}_{{\bm n}_k,
{\bm n}_j}^{1/N} \right)^N | {\bm n}_j \rangle \Big]
\nonumber \\
&=& \mathrm{arg} \left[ \langle {\bm n}_k | \tilde{U}_{{\bm n}_k,
{\bm n}_{kj}^{(N-1)}} \ldots \tilde{U}_{{\bm n}_{kj}^{(2)}, {\bm
n}_{kj}^{(1)}} \tilde{U}_{{\bm n}_{kj}^{(1)}, {\bm n}_j} | {\bm n}_j
\rangle \right],
\nonumber \\
\label{eq-inv-gamma-2}
\end{eqnarray}
where $N \rightarrow \infty$ is a large number, and ${\bm
n}_{kj}^{(n)}$ in terms of $n = 1, 2, \ldots, N-1$ are unit vectors
equally spaced along the great circle connecting ${\bm n}_j$ and
${\bm n}_k$ [see Fig.~\ref{geometry}~(a)]. At the intermediate step between the infinitesimal spin
rotations $\tilde{U}_{{\bm n}_{kj}^{(n)}, {\bm n}_{kj}^{(n-1)}}$ and
$\tilde{U}_{{\bm n}_{kj}^{(n+1)}, {\bm n}_{kj}^{(n)}}$ in
Eq.~(\ref{eq-inv-gamma-2}), the spin state is proportional to $|
{\bm n}_{kj}^{(n)} \rangle$ by construction. If we then use the
orthogonality relation $\langle -{\bm p} | {\bm p} \rangle = 0$ and
the resolution of identity
\begin{equation}
| {\bm p} \rangle \langle {\bm p} | + | {-}{\bm p} \rangle \langle
-{\bm p} | = 1 \label{eq-inv-res}
\end{equation}
for $| {\bm p} \rangle = | {\bm n}_{kj}^{(n)} \rangle$ at each
intermediate step, and also Eq.~(\ref{eq-rot-qp}) to turn each
geodesic matrix element back into an overlap, the Berry connection
in Eq.~(\ref{eq-inv-gamma-2}) becomes
\begin{widetext}
\begin{eqnarray}
\gamma_{kj} &=& \mathrm{arg} \left[ \langle {\bm n}_k |
\tilde{U}_{{\bm n}_k, {\bm n}_{kj}^{(N-1)}} | {\bm n}_{kj}^{(N-1)}
\rangle \ldots \langle {\bm n}_{kj}^{(2)} | \tilde{U}_{{\bm
n}_{kj}^{(2)}, {\bm n}_{kj}^{(1)}} | {\bm n}_{kj}^{(1)} \rangle
\langle {\bm n}_{kj}^{(1)} | \tilde{U}_{{\bm n}_{kj}^{(1)}, {\bm
n}_j} | {\bm n}_j \rangle \right]
\nonumber \\
&=& \mathrm{arg} \left[ \langle {\bm n}_k | {\bm n}_{kj}^{(N-1)}
\rangle \right] + \ldots + \mathrm{arg} \left[ \langle {\bm
n}_{kj}^{(2)} | {\bm n}_{kj}^{(1)} \rangle \right] + \mathrm{arg}
\left[ \langle {\bm n}_{kj}^{(1)} | {\bm n}_j \rangle \right].
\label{eq-inv-gamma-3}
\end{eqnarray}
\end{widetext}
Consequently, the Berry phase in Eq.~(\ref{eq-gen-gamma-1}) is a sum
of infinitely many infinitesimal Berry connections along a closed
loop in the Hilbert space and can thus be converted into an
appropriate integral of the Berry curvature along a surface bounded
by this closed loop. For spin states, it is well known that the
Berry curvature is $1/2$ everywhere on the Bloch sphere, and the
Berry phase $\gamma_{jkl}$ in
Eq.~(\ref{eq-gen-gamma-1}) is therefore half the solid angle of the
spherical triangle spanned by ${\bm n}_j$, ${\bm n}_k$, and ${\bm
n}_l$.

\begin{figure*}[!htb]
\includegraphics[width=18cm]{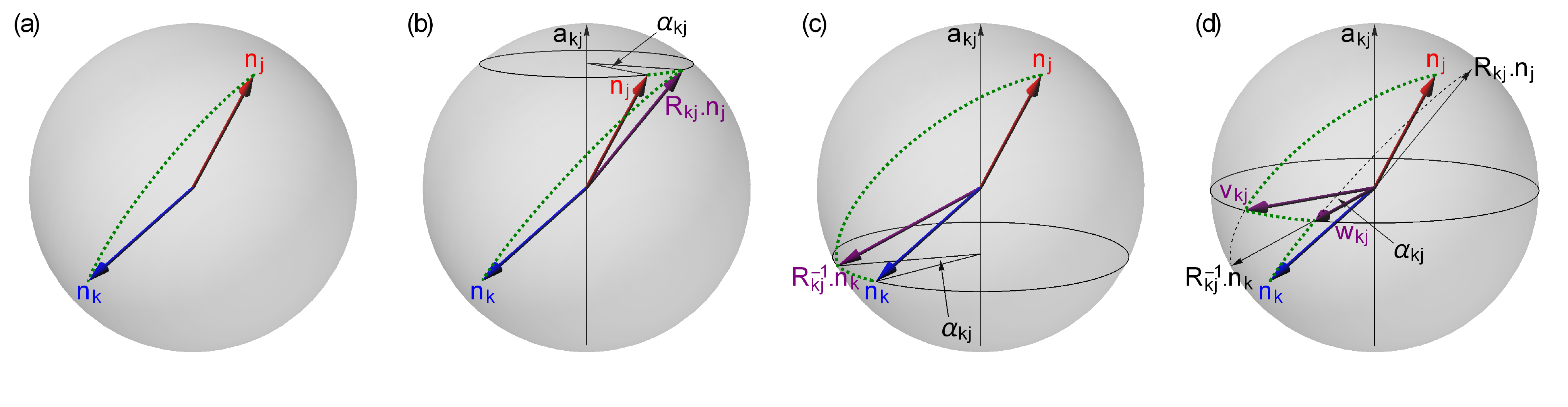}
\caption{(a) Great circle connecting the unit
vectors ${\bm n}_j$ and ${\bm n}_k$. (b) The spin direction is
first rotated around the axis ${\bm a}_{kj}$ and
then along the great circle containing $R_{kj} \cdot {\bm n}_j$ and
${\bm n}_k$. (c) The spin direction is first rotated along the great
circle containing ${\bm n}_j$ and $R^{-1}_{kj}
\cdot {\bm n}_k$ and then around the axis ${\bm a}_{kj}$. (d)
Unit vectors ${\bm v}_{kj}$ and ${\bm w}_{kj}$ along
the great circle perpendicular to ${\bm a}_{kj}$. Using these two
intermediate unit vectors, the spin rotation $U_{kj}$ becomes a
geodesic spin rotation $\tilde{U}_{{\bm w}_{kj}, {\bm v}_{kj}}$.}
\label{geometry}
\end{figure*}

\subsection{General case with spin-orbit interaction} \label{sec-soc}

To calculate the general Berry connection in
Eq.~(\ref{eq-gen-gamma-2}), we first notice that $U_{kj} | {\bm n}_j
\rangle \propto | R_{kj} \cdot {\bm n}_j \rangle$.
Using Eq.~(\ref{eq-inv-res}) for $| {\bm p} \rangle = | R_{kj} \cdot
{\bm n}_j \rangle$, the Berry connection in
Eq.~(\ref{eq-gen-gamma-2}) then becomes
\begin{equation}
\gamma_{kj} = \mathrm{arg} \Big[ \langle {\bm n}_k | R_{kj} \cdot
{\bm n}_j \rangle \langle R_{kj} \cdot {\bm n}_j | U_{kj} |
{\bm n}_j \rangle \Big]. \label{eq-soc-gamma-1}
\end{equation}
Next, if we employ Eq.~(\ref{eq-rot-qp}) to turn the overlap into a
geodesic matrix element, and use Eq.~(\ref{eq-inv-res}) with $| {\bm
p} \rangle = | R_{kj} \cdot {\bm n}_j \rangle$ again, the Berry
connection takes the form
\begin{eqnarray}
\gamma_{kj} &=& \mathrm{arg} \Big[ \langle {\bm n}_k |
\tilde{U}_{{\bm n}_k, R_{kj} \cdot {\bm n}_j} | R_{kj} \cdot
{\bm n}_j \rangle \langle R_{kj} \cdot {\bm n}_j | U_{kj} |
{\bm n}_j \rangle \Big] \nonumber \\
&=& \mathrm{arg} \Big[ \langle {\bm n}_k | \tilde{U}_{{\bm n}_k,
R_{kj} \cdot {\bm n}_j} U_{kj} | {\bm n}_j \rangle \Big].
\label{eq-soc-gamma-2}
\end{eqnarray}
In this representation, the spin direction is first rotated around
the axis ${\bm a}_{kj}$ and then along the great circle containing
$R_{kj} \cdot {\bm n}_j$ and ${\bm n}_k$ [see Fig.~\ref{geometry}~(b)]. By exploiting the identity
\begin{equation}
U_{kj}^{-1} \tilde{U}_{{\bm n}_k, R_{kj} \cdot {\bm n}_j} U_{kj} =
\tilde{U}_{R_{kj}^{-1} \cdot {\bm n}_k, {\bm n}_j},
\label{eq-soc-U-1}
\end{equation}
we can also obtain an alternative formula for the Berry connection:
\begin{equation}
\gamma_{kj} = \mathrm{arg} \Big[ \langle {\bm n}_k | U_{kj}
\tilde{U}_{R_{kj}^{-1} \cdot {\bm n}_k, {\bm n}_j} | {\bm n}_j
\rangle \Big]. \label{eq-soc-gamma-3}
\end{equation}
In this representation, the spin direction is first rotated along
the great circle containing ${\bm n}_j$ and $R_{kj}^{-1} \cdot {\bm
n}_k$ and then around the axis ${\bm a}_{kj}$ [see
Fig.~\ref{geometry}~(c)]. Generically, in either of these two
representations, the rotation around the axis ${\bm
a}_{kj}$ is \emph{not} along a great circle of the Bloch sphere.
However, there is a ``mixed'' representation between these two
``pure'' representations in which the spin direction is rotated
along great circles all the way from ${\bm n}_j$ to ${\bm n}_k$:
first along the one containing ${\bm n}_j$ and $R_{kj}^{-1} \cdot
{\bm n}_k$, then along the one perpendicular to
${\bm a}_{kj}$, and finally along the one
containing $R_{kj} \cdot {\bm n}_j$ and ${\bm n}_k$. To find the
corresponding formula for the Berry connection, we start from
Eq.~(\ref{eq-soc-gamma-3}) and notice that there is a general
identity
\begin{eqnarray}
U_{kj} \tilde{U}_{R_{kj}^{-1} \cdot {\bm n}_k, {\bm n}_j} &=& \pm
U_{kj} \tilde{U}_{R_{kj}^{-1} \cdot {\bm n}_k, {\bm v}_{kj}}
\tilde{U}_{{\bm v}_{kj}, {\bm n}_j}
\nonumber \\
&=& \pm \tilde{U}_{{\bm n}_k,
R_{kj} \cdot {\bm v}_{kj}} U_{kj} \tilde{U}_{{\bm v}_{kj},
{\bm n}_j} \label{eq-soc-U-2}
\end{eqnarray}
for any unit vector ${\bm v}_{kj}$ along the great circle containing
${\bm n}_j$ and $R_{kj}^{-1} \cdot {\bm n}_k$. The sign ambiguity in
Eq.~(\ref{eq-soc-U-2}) reflects that the left-hand side and the
right-hand side may differ in a $2\pi$ spin rotation $e^{i \pi} =
-1$. We ignore this sign ambiguity as it does not matter for our
purposes. Next, if the unit vector ${\bm v}_{kj}$ is also along the
great circle perpendicular to ${\bm a}_{kj}$ [see Fig.~\ref{geometry}~(d)], the spin rotation
$U_{kj}$ becomes a geodesic spin rotation $\tilde{U}_{{\bm w}_{kj},
{\bm v}_{kj}}$, where ${\bm w}_{kj} = R_{kj} \cdot {\bm v}_{kj}$,
and thus Eq.~(\ref{eq-soc-U-2}) takes the form
\begin{equation}
U_{kj} \tilde{U}_{R_{kj}^{-1} \cdot {\bm n}_k, {\bm n}_j} = \pm
\tilde{U}_{{\bm n}_k, {\bm w}_{kj}} \tilde{U}_{{\bm w}_{kj},
{\bm v}_{kj}} \tilde{U}_{{\bm v}_{kj}, {\bm n}_j}.
\label{eq-soc-U-3}
\end{equation}
The unit vector ${\bm v}_{kj}$ is then along the intersection of the
great circle perpendicular to ${\bm a}_{kj}$ and
the great circle containing ${\bm n}_j$ and $R_{kj}^{-1} \cdot {\bm
n}_k$:
\begin{equation}
{\bm v}_{kj} = \frac{{\bm a}_{kj} \times \left[ {\bm n}_j \times
\left( R_{kj}^{-1} \cdot {\bm n}_k \right) \right]} {\left| {\bm
a}_{kj} \times \left[ {\bm n}_j \times \left( R_{kj}^{-1} \cdot {\bm
n}_k \right) \right] \right|}, \label{eq-soc-v}
\end{equation}
while the unit vector ${\bm w}_{kj}$ is along the intersection of
the great circle perpendicular to ${\bm a}_{kj}$
and the great circle containing $R_{kj} \cdot {\bm n}_j$ and ${\bm
n}_k$:
\begin{equation}
{\bm w}_{kj} = R_{kj} \cdot {\bm v}_{kj} = \frac{{\bm a}_{kj} \times
\left[ \left( R_{kj} \cdot {\bm n}_j \right) \times {\bm n}_k
\right]} {\left| {\bm a}_{kj} \times \left[ \left( R_{kj} \cdot {\bm
n}_j \right) \times {\bm n}_k \right] \right|}. \label{eq-soc-w}
\end{equation}
Note that $-{\bm v}_{kj}$ and $-{\bm w}_{kj}$ are also along the
intersections of the same great circles and could thus be used
instead of ${\bm v}_{kj}$ and ${\bm w}_{kj}$ in
Eq.~(\ref{eq-soc-U-3}). Substituting Eq.~(\ref{eq-soc-U-3}) into
Eq.~(\ref{eq-soc-gamma-3}), and using Eqs.~(\ref{eq-rot-qp}) and
(\ref{eq-inv-res}), the Berry connection finally becomes
\begin{widetext}
\begin{eqnarray}
\gamma_{kj} &=& \mathrm{arg} \Big[ {\pm} \langle {\bm n}_k |
\tilde{U}_{{\bm n}_k, {\bm w}_{kj}} \tilde{U}_{{\bm w}_{kj},
{\bm v}_{kj}} \tilde{U}_{{\bm v}_{kj}, {\bm n}_j} | {\bm n}_j
\rangle \Big]
= \mathrm{arg} \Big[ {\pm} \langle {\bm n}_k |
\tilde{U}_{{\bm n}_k, {\bm w}_{kj}} | {\bm w}_{kj} \rangle \langle
{\bm w}_{kj} | \tilde{U}_{{\bm w}_{kj}, {\bm v}_{kj}} | {\bm v}_{kj}
\rangle \langle {\bm v}_{kj} | \tilde{U}_{{\bm v}_{kj}, {\bm n}_j} |
{\bm n}_j \rangle \Big]
\nonumber  \\
&=& \mathrm{arg} \left[ \langle {\bm n}_k | {\bm w}_{kj} \rangle
\right] + \mathrm{arg} \left[ \langle {\bm w}_{kj} | {\bm v}_{kj}
\rangle \right] + \mathrm{arg} \left[ \langle {\bm v}_{kj} |
{\bm n}_j \rangle \right] \quad (\mathrm{mod} \, \pi). \nonumber
\end{eqnarray}
\end{widetext}
Consequently, the Berry phase is a sum of nine Berry connections,
each taking the simplified form of Eq.~(\ref{eq-inv-gamma-1}).
Repeating the steps of Sec.~\ref{sec-inv}, the Berry phase
$\gamma_{jkl}$ in Eq.~(\ref{eq-gen-gamma-1}) is
then half the solid angle of the spherical \emph{nonagon} spanned by
${\bm n}_j$, ${\bm v}_{kj}$, ${\bm w}_{kj}$, ${\bm n}_k$, ${\bm
v}_{lk}$, ${\bm w}_{lk}$, ${\bm n}_l$, ${\bm v}_{jl}$, and ${\bm
w}_{jl}$. While this result is only valid modulo $\pi$, it can be
used to deduce if the effective flux produced by the combination of
magnetic ordering and SOC breaks the time-reversal symmetry or not.

The most important consequence of this result is that the Berry
curvature (i.e., the effective magnetic field) can be non-zero even
for \emph{collinear} or \emph{coplanar} spin configurations if the
spin-orbit interaction is finite. This feature will become clearer
in the next section, where we solve the same problem by using a
convenient rotation of the local reference frame at each individual
spin.

\section{Algebraic approach}
\label{sec-alg}

\subsection{SU(2) invariant case}

We have seen in the previous section that, in absence of spin-orbit
coupling, the Berry phase $\gamma_{jkl}$ picked up by an electron as
it moves around a triangle of spins $jkl$ is half the solid angle
subtended by the three spins:
\begin{equation}
\gamma_{jkl}=\frac{\Omega_{jkl}}{2}.
\label{Berrysu2}
\end{equation}
This simple equation can be rederived in the following way. We first
introduce an arbitrary unit vector ${\bm{n}}$ that we choose as our
quantization axis, i.e., ${\bm{n}}={\hat{z}}$. In this frame, the
states of the spins $j$ and $k$ are:
\begin{eqnarray}
|{\bm n}_{j}\rangle & = & \cos{\frac{\theta_{j}}{2}}|\uparrow\,\rangle+e^{i\phi_{j}}\sin{\frac{\theta_{j}}{2}}|\downarrow\,\rangle,\nonumber \\
|{\bm n}_{k}\rangle & = &
\cos{\frac{\theta_{k}}{2}}|\uparrow\,\rangle+e^{i\phi_{k}}\sin{\frac{\theta_{k}}{2}}|\downarrow\,\rangle.
\end{eqnarray}
Given that $\phi_{j}-\phi_{k}$ is defined modulo $2\pi$, we will use
this freedom to require that $|\phi_{j}-\phi_{k}|\leq\pi$.  In
absence of SOC, the Berry connection becomes
\begin{eqnarray}
\gamma_{kj} &=& \arg [\langle {\bm n}_{k}  | {\bm n}_{j}\rangle] \\
&=& \arg \left[
\cos{\frac{\theta_{j}}{2}}\cos{\frac{\theta_{k}}{2}}+e^{i(\phi_{j}-\phi_{k})}\sin{\frac{\theta_{j}}{2}}\sin{\frac{\theta_{k}}{2}}
\right], \nonumber
\end{eqnarray}
implying that
\begin{eqnarray}
\tan{\gamma_{kj}} &=&
\frac{\sin{(\phi_{j}-\phi_{k})}}{\cot{\frac{\theta_{j}}{2}}\cot{\frac{\theta_{k}}{2}}
+ \cos{(\phi_{j}-\phi_{k})}}
\nonumber \\
&=&
\tan{\frac{\Omega(\phi_{j}-\phi_{k},\theta_{j},\theta_{k})}{2}},\label{solidangle}
\end{eqnarray}
where
$\Omega(\phi_{j}-\phi_{k},\theta_{j},\theta_{k})$
is the solid angle subtended by the vectors $({\bm{n}},
{\bm{n}}_{j}, {\bm{n}}_{k})$. Since the quantization axis ${\bm{n}}$
is the same for the three bonds of the triangle, the Berry phase is
indeed given by Eq.~\eqref{Berrysu2}:
\begin{equation}
\gamma_{jkl}=\gamma_{kj}+\gamma_{lk}+\gamma_{jl}=\frac{\Omega_{jkl}}{2},\label{Berrysu2b}
\end{equation}
where $\Omega_{jkl}$ is the solid angle subtended by the vectors
$({\bm{n}}_{j}, {\bm{n}}_{k}, {\bm{n}}_{l})$ corresponding to the
spin directions.

\subsection{General case with spin-orbit interaction}

Our next goal is to generalize Eq.~\eqref{Berrysu2b} for the case of
finite SOC, where:
\begin{equation}
\gamma_{jkl} = \mathrm{arg} \left[ \langle {\bm n}_k | U_{kj} | {\bm
n}_j \rangle \langle {\bm n}_l | U_{lk} | {\bm n}_k \rangle \langle
{\bm n}_j | U_{jl} | {\bm n}_l \rangle \right], \label{BerrySOC}
\end{equation}
We first notice that the Berry phase $\gamma_{jkl}$ is invariant
under local rotations of the spin reference frame,
\begin{eqnarray}
| {\bm n}'_j \rangle &=& {\cal U}_{j} |  {{\bm n}}_j \rangle, \;\;\;
|   {\bm n}'_k \rangle =  {\cal U}_{k} | {{\bm n}}_k \rangle, \;\; |
{\bm n}'_l \rangle =  {\cal U}_{l} | {{\bm n}}_l \rangle,
\label{lsre} \\
{\bm n}'_j &=& {\cal R}_j \cdot {\bm n}_j, \;\;\; {\bm n}'_k = {\cal
R}_k \cdot {\bm n}_k, \;\;\; {\bm n}'_l =  {\cal R}_l \cdot {\bm
n}_l, \label{lsrl}
\end{eqnarray}
where ${\cal R}_j$ is the SO(3) rotation matrix associated with
the SU(2) matrix ${\cal U}_{j}$, if we
simultaneously transform the unitary operator on each bond  in the
following way:
\begin{equation}
U'_{kj}  =  {\cal U}_{k}  U_{kj}  {\cal U}^{\dagger}_{j}, \;\;\;
U'_{lk}  =  {\cal U}_{l}  U_{lk}  {\cal U}^{\dagger}_{k}, \;\;\;
U'_{jl}  =  {\cal U}_{j}  U_{jl}  {\cal U}^{\dagger}_{l}.
\label{guu}
\end{equation}
This observation simply reflects the gauge invariance of the Berry
phase under rotations of the  local spin spin reference frame and it motivates the introduction
of the Wilson loop operator
\begin{equation}
{\cal A}_{jkl} = U_{jl} U_{lk}  U_{kj} = \exp \left[ -\frac{i
\alpha_{jkl}} {2} ({\bm{a}}_{jkl} \cdot {\boldsymbol \sigma})
\right], \label{Wilson-loop}
\end{equation}
which is also a gauge invariant quantity. We note that the exchange
Hamiltonian ${\cal H}_J$ between the spins of the
itinerant electrons and the local magnetic moments remains invariant
under the local spin rotations described by Eqs.~\eqref{lsre} and
\eqref{lsrl}.

The next step is to perform a convenient rotation of the local
reference frame of two spins (say $k$ and $l$) such that the unitary
operator becomes the identity on two out of three bonds (say $kj$
and $jl$). To this end, one can use the local unitary
transformations
\begin{eqnarray}
{\cal U}_{j} &=&  I, \;\;\;
{\cal U}_{k} = U^{\dagger}_{kj}, \;\;\;
{\cal U}_{l} = U_{jl},
\nonumber \\
{\cal R}_{j} &=&  I, \;\;\; {\cal R}_{k} = R^{T}_{kj}, \;\;\; {\cal
R}_{l} = R_{jl}.
\end{eqnarray}
Given that the Wilson loop remains invariant under such a
transformation, the unitary operator on the third bond $lk$ must
then be equal to the Wilson loop ${\cal A}_{jkl}$.

The final step is to align the global quantization axis ${\bm n}$
with the rotation axis of the Wilson loop operator: ${\bm{n}} =
{\bm{a}}_{jkl}$. The Berry connections can then be easily computed
in the new reference frame:
\begin{widetext}
\begin{eqnarray}
\gamma_{kj} &=& \arg{[\langle {\bm n}'_k| {\bm
n}'_j\rangle]} = \frac{\Omega(\phi'_j - \phi'_k,
\theta'_j, \theta'_k)}{2},
\nonumber \\
\gamma_{jl} &=& \arg{[\langle {\bm n}'_j| {\bm
n}'_l\rangle]} = \frac{\Omega(\phi'_l - \phi'_j,
\theta'_l, \theta'_j)}{2},
\\
\gamma_{lk} &=& \arg{[\langle {\bm n}'_l|{\cal
A}_{jkl}| {\bm n}'_k\rangle]} = \arg \bigg[
e^{-\frac{i}{2}\alpha_{jkl}} \cos{\frac{ \theta'_{k}}{2}}
\cos{\frac{ \theta'_{l}}{2}} + e^{i( \phi'_{k}-
\phi'_{l}+\frac{1}{2}\alpha_{jkl})} \sin{\frac{
\theta'_{k}}{2}} \sin{\frac{\theta'_{l}}{2}} \bigg]
\nonumber \\
&=& -\frac{\alpha_{jkl}}{2} + \frac{ \Omega(  \phi'_k -
{\phi}'_l + \alpha_{jkl},  \theta'_k,
\theta'_l)}{2}. \nonumber
\end{eqnarray}
\end{widetext}
Thus, the Berry phase picked up by an electron as it moves around
the triangle $jkl$ is
\begin{equation}
\gamma_{jkl} = \frac{\Omega'_{jkl}}{2} + {\hat
\gamma}_{jkl}, \label{Berrysol}
\end{equation}
where $\Omega'_{jkl}$ is the solid angle subtended by the
three vectors $({\bm n}'_j, {\bm n}'_k, {\bm
n}'_l)$ corresponding to the rotated spin directions in the new
reference frame, and
\begin{equation}
{\hat \gamma}_{jkl} = -\frac{\alpha_{jkl}}{2} + \frac{\delta \Omega'_{jkl}}{2}, \label{Berrysol-hat}
\end{equation}
in terms of
\begin{equation}
\delta  \Omega'_{jkl} = \Omega( \phi'_k -
\phi'_l + \alpha_{jkl}, \theta'_k, \theta'_l) -
\Omega(\phi'_k - \phi'_l,  \theta'_k,
\theta'_l), \label{delomegajkl}
\end{equation}
is an additional contribution due to the Wilson loop ${\cal
A}_{jkl}$.

Eq.~\eqref{Berrysol} generalizes Eq.~\eqref{Berrysu2}, which is only
valid for SU(2) invariant systems. In particular, it is easy to
demonstrate that collinear or coplanar configurations can induce a
Berry phase different from $0$ or $\pi$, which acts
as an effective magnetic flux. As an example, we can consider the
case of a collinear ferromagnet, ${\bm n}_{j} \parallel {\bm n}_{k}
\parallel{\bm n}_{j}$ with the three vectors ${\bm a}_{kj}$, ${\bm
a}_{lk}$, and ${\bm a}_{jl}$ being  parallel (or antiparallel) to
the magnetization. In this case, Eq.~\eqref{Berrysol} tells us that
that $\gamma_{jkl} = -\alpha_{jkl}/2$, implying
that the combination of ferromagnetism and SOC  generates
a real space Berry phase that is
proportional to the rotation angle of the Wilson loop. In this way,
we recover the essential result of Karplus and
Luttinger~\cite{Karplus54} in the minimal model that we are
considering here. The potential emergence of real space Berry
curvature in coplanar antiferromagnets will become clearer in the
next section, where we discuss the limit of small SOC. However, one
can immediately verify that Eq.~\eqref{Berrysol} also gives a finite
Berry curvature for the coplanar spin configuration illustrated in
Figure~\ref{triang}.

\begin{figure}[!htb]
 \includegraphics[width=5.5cm]{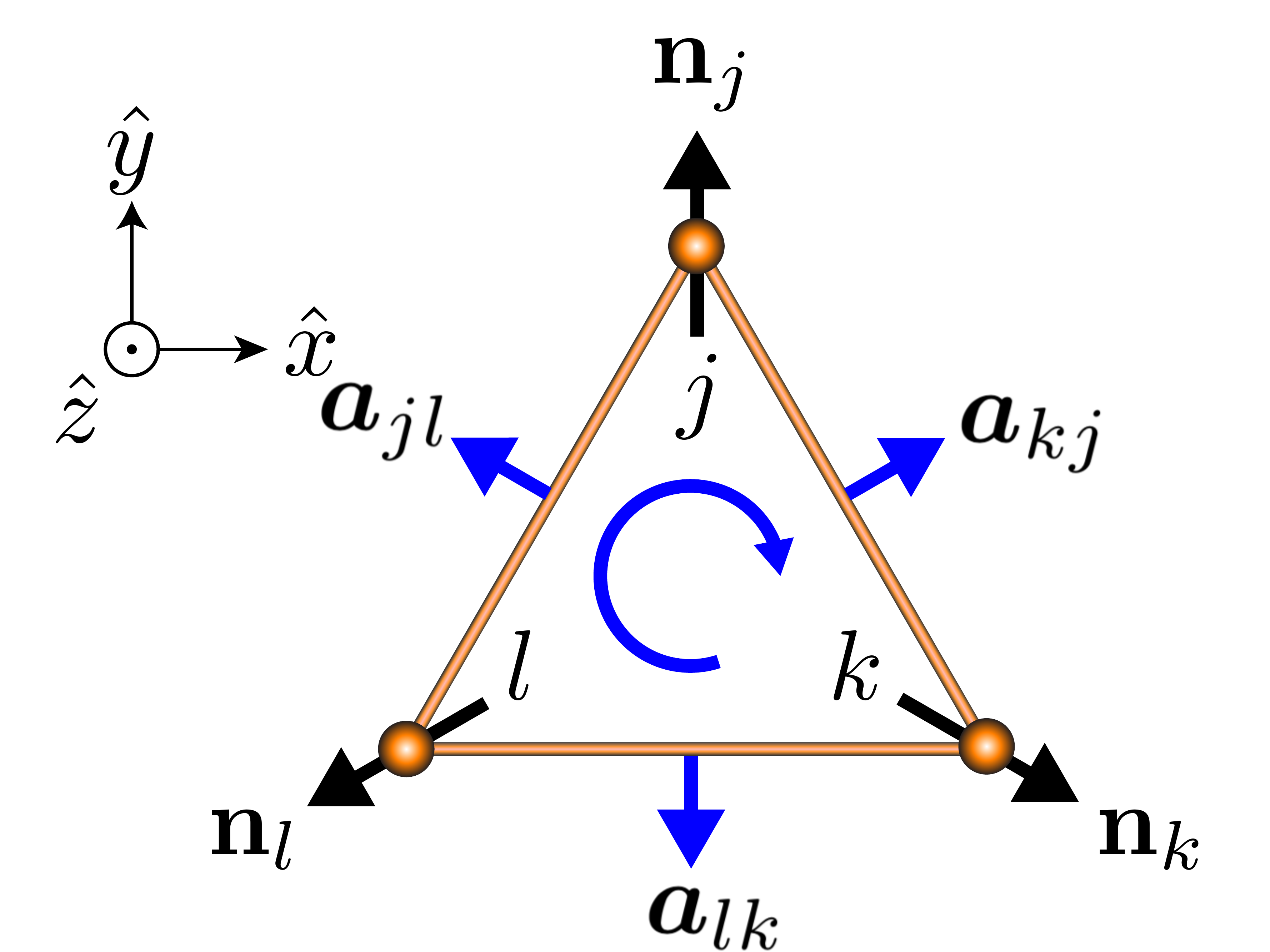}
 \caption{Coplanar configuration with Berry phase different from 0 or $\pi$.
The clockwise circulation indicates the bond
orientations required to define the directions of
the spin-orbit vectors ${\bm a}_{kj}$, ${\bm
a}_{lk}$, and ${\bm a}_{jl}$, indicated with blue arrows.} \label{triang}
\end{figure}

\subsection{Limit of small spin-orbit interaction}

We will now consider the quite general case of small spin-orbit
coupling: $\alpha_{kj} \ll 1$ for all bond bonds $kj$ connected by
finite hopping amplitudes $t_{kj}$ (the hopping amplitudes are
assumed to be zero beyond a characteristic distance of a few lattice
spaces because of the exponential decay of the atomic orbitals).
Expanding Eqs.~\eqref{eq-gen-U} and \eqref{Wilson-loop} up to first
order in $\alpha_{kj}$ and $\alpha_{jkl}$, respectively, the
parameters of the Wilson loop operator then easily follow from those
of the individual unitary operators:
\begin{eqnarray}
\alpha_{jkl} &=& |{\bm{v}}_{jkl}|, \quad\,\, {\bm{a}}_{jkl} =
\frac{{\bm{v}}_{jkl}}{|{\bm{v}}_{jkl}|},
\nonumber \\
{\bm{v}}_{jkl} &=&
\alpha_{kj}{\bm{a}}_{kj}+\alpha_{lk}{\bm{a}}_{lk}+\alpha_{jl}{\bm{a}}_{jl}.
\end{eqnarray}
Moreover, the contribution to the Berry phase in
Eq.~\eqref{Berrysol-hat} can be expanded up to first order in
$\alpha_{jkl}$ as
\begin{equation}
{\hat \gamma}_{jkl} = \frac{\alpha_{jkl}}{2} \left[ -1 +
\frac{\partial \Omega( \phi'_k - \phi'_l + \alpha,
 \theta'_k, \theta'_l)} {\partial \alpha}
\bigg|_{\alpha = 0} \right],
\end{equation}
and can be brought to a simple form via Eq.~\eqref{solidangle}:
\begin{equation}
{\hat \gamma}_{jkl} = -\frac{ \alpha_{jkl} (\cos{ \theta'_k} +
\cos { \theta'_l)} } {2 [1 + \cos { \theta}'_k \cos {
\theta}'_l + \sin { \theta}'_k \sin { \theta}'_l \cos
({ \phi}'_k - { \phi}'_l)]}.
\end{equation}
In particular, when the spin directions are close to a ferromagnetic
configuration, such that ${\theta}'_k \approx {
\theta}'_l \approx {\theta}'$ and ${\phi}'_k \approx
{ \phi}'_l \approx {\phi}'$, this contribution becomes
\begin{equation}
{\hat \gamma}_{jkl} = -\frac{\alpha_{jkl} \cos {\theta'}} {2},
\label{llwll}
\end{equation}
which is simply {\it the projection of the net spin-orbit rotation to the
common direction of the spins}.

In the special case when the Wilson loop is equal to the identity
($\alpha_{jkl} = 0$), the Berry phase ${\hat \gamma}_{jkl}$ in
Eq.~\eqref{Berrysol-hat} vanishes. The total Berry phase
$\gamma_{jkl}$ in Eq.~\eqref{Berrysol} is then half the solid angle
${\Omega}'_{jkl}$ subtended by the three magnetic moments {\it
in the local reference frame} that is required to ``gauge away'' the
SOC. Clearly, even in this case, an
antiferromagnetic ordering that is coplanar in the original
reference frame can be non-coplanar in the rotated reference frame
and thus produce a finite real-space Berry curvature.

For example, in the C$_{3}$ invariant system depicted in
Figure~\ref{triang}, the three angles, $\alpha_{kj} = \alpha_{lk} =
\alpha_{jl} = \alpha \ll 1$, are identical, and the three vectors,
${\bm{a}}_{kj}$, ${\bm{a}}_{lk}$, and ${\bm{a}}_{jl}$, are related
by $2\pi/3$ rotations around the ${\hat {\bm z}}$ axis.
Consequently,
\begin{equation}
{\bm{v}}_{jkl} = \alpha({\bm{a}}_{kj} + {\bm{a}}_{lk} +
{\bm{a}}_{jl}) = 0
\end{equation}
if the $z$-component of ${\bm{a}}_{kj}$ vanishes. In other words, to
first order in the SOC, the Wilson loop is equal to the identity:
${\cal A}_{jkl}=I$. To the same order, the three spin directions in
the new reference frame are
\begin{equation}
{\bm n}'_j = {\bm n}_j, \; {\bm n}'_k = {\bm n}_k -
\alpha {\bm a}_{kj} \times {\bm n}_{k}, \; {\bm n}'_l = {\bm
n}_l + \alpha {\bm a}_{jl} \times {\bm n}_{l},
\end{equation}
and the Berry phase is thus given by
\begin{eqnarray}
\gamma_{jkl} &=& \frac{\Omega'_{jkl}}{2} = \pi - 6 \alpha +
{\cal O}(\alpha^2).
\end{eqnarray}
The simple message of this example is that, whenever the SOC can be
gauged away in a particular local reference frame (i.e., the Wilson
loop is equal to the identity), {\it the magnetic ordering must be
non-coplanar in that reference frame to produce a finite real-space
Berry curvature}. It is easy to imagine that antiferromagnetic
orderings that are collinear or coplanar in the original reference
frame can become non-coplanar in the rotated reference frame leading
to a finite Berry curvature.

\section{Continuum limit}
\label{sec-cont}

The simple ideas that we discussed in the previous section  can be
presented in a more formal and elegant way by taking the continuum
limit. This limit is appropriate for describing situations where the
SU(2) Wilson loop  bond field defined by ${\cal A}_{jkl}$,
associated with the SOC, and the magnetic texture vary over a length
scale which is much longer than the lattice parameter. The first
condition can be realized by long wavelength
lattice deformations induced by strain.  The second condition arises
naturally in materials  with a very small magnetic ordering wave
vector.

By taking the continuum limit, we will find an explicitly covariant
form of the effective magnetic field or real space Berry curvature
produced by the underlying vector field ${\bm n}_j$ in presence of
SOC. We have seen that  the  unitary ``hopping'' matrices ${U}_{jk}$
correspond to a non-Abelian SU(2) gauge field which
is fixed by the interplay between the lattice structure and the
SOC.~\footnote{This field becomes a dynamical  variable if the ionic
positions are allowed to fluctuate.} After taking the continuum
limit,  the bond matrices  ${U}_{jk}$ become infinitesimal SU(2)
rotations connecting the points ${\bm x}$ and ${\bm x} + d {\bm x}$
which are parametrized by the field $A^a_{\mu}$:
\begin{equation}
U_{{\bm x} + d {\bm x},{\bm x}} = \exp{\left [\frac{1}{2} \sigma^a A^a_{\mu} dx_{\mu}  \right]},
\end{equation}
where repeated indices are implicitly summed over.

Once again, we want to compute the Berry phase that the electronic
wave function acquires when the electron moves along a one
dimensional closed path  ${\cal C}= \{ x_{\mu}(\tau) \; {\rm with}
\; \tau \in [0; T], x_{\mu}(0)= x_{\mu}(T) \}$. As we did for the
lattice case, we will first derive the well-known expression of the
Berry phase in the SU(2) invariant case and we will use this result
as an introduction for obtaining a gauge invariant form of the Berry
curvature or effective magnetic field in the presence of SOC.

\subsection{SU(2) invariant case}

For the SU(2) invariant case on the lattice, we have seen that the
Berry phase for a closed loop ${\cal
C}:\;j\rightarrow k\rightarrow l\rightarrow...\rightarrow m
\rightarrow j$ is:
\begin{equation}
\gamma_{\cal C}  =  \text{arg}\left[\langle\bm{n}_{j}
\rvert\bm{n}_{m}\rangle...\langle\bm{n}_{l}
\rvert\bm{n}_{k}\rangle\langle\bm{n}_{k}
\rvert\bm{n}_{j}\rangle\right]. \label{elem-loop}
\end{equation}
Note that an arbitrary closed loop  can be obtained from a superposition of multiple ``elementary'' triangular loops.
To find the counterpart of Eq.~\eqref{elem-loop} in the continuum,  we will divide the interval $T$ into $N$ equal subintervals $\Delta \tau = T/N $ to finally take the $N \to \infty$ limit. In this way we obtain:
\begin{eqnarray}
\gamma_{\cal C} &=& \lim_{N \to \infty} \mathrm{arg} \left \{
\prod_{j=0}^{N-1} \langle {\bm n}_{{\bm x}(\tau_j+\Delta \tau  )}  |
{\bm n}_{{\bm x}(\tau_j)} \rangle \right\}
\nonumber \\
&=& \lim_{N \to \infty} \mathrm{arg} \left \{  \prod_{j=0}^{N-1} [1
- \langle {\bm n}_{{\bm x}(\tau_j)} | \partial_{\tau} |{\bm n}_{{\bm
x}(\tau_j)} \rangle  \Delta \tau] \right \}
\nonumber \\
&=& \lim_{N \to \infty} \mathrm{arg} \left \{ \prod_{j=0}^{N-1}
\exp{  [ - \langle {\bm n}_{{\bm x}(\tau_j)} | \partial_{\tau}| {\bm
n}_{{\bm x}(\tau_j)} \rangle \Delta \tau ] } \right \}
\nonumber \\
&=& i \lim_{N \to \infty} \sum_{j=0}^{N-1} \langle {\bm n}_{{\bm
x}(\tau_j)} | \partial_{\tau}| {\bm n}_{{\bm x}(\tau_j)} \rangle
\Delta \tau
\nonumber \\
&=&  i \int_0^{T}  \langle {\bm n}_{{\bm x}(\tau)} |
\partial_{\tau}| {\bm n}_{{\bm x}(\tau)} \rangle d\tau,
\label{Berry-for}
\end{eqnarray}
where $\tau_j = j \Delta \tau$. The geometric character of the Berry phase becomes evident after reexpressing  Eq.~\eqref{Berry-for} in terms of a closed integral over the loop ${\cal C}$:
\begin{equation}
\gamma_{\cal C}  = i \oint_{{\cal C}}
\langle\bm{n}\rvert\partial_{\mu}\rvert\bm{n}\rangle dx_{\mu}  =
\frac{1}{2}\int_{S_{{\cal
C}}}\bm{n}\cdot[\partial_{\mu}\bm{n}\times\partial_{\nu}\bm{n}]d^{2}\sigma^{\mu\nu}
= \frac{\Omega_{\cal C}}{2},\label{eq:berry_curvature}
\end{equation}
where $S_{{\cal C}}$ is the area enclosed by the
loop ${\cal C}$, while $\Omega_{\cal C}$ is the solid angle
subtended by the vector field ${\bm n}$ around the
loop ${\cal C}$. This equation corresponds to the continuum limit of
Eq.~\eqref{Berrysu2}.

\subsection{General case with spin-orbit interaction}

For the lattice problem with finite SOC, Eq.~\eqref{elem-loop} must be generalized to:
\begin{equation}
\gamma_{{\cal C}} = \text{arg}\left[\langle\bm{n}_{j}\rvert
U_{jm}\rvert\bm{n}_{m}\rangle...\langle\bm{n}_{l}\rvert
U_{lk}\rvert\bm{n}_{k}\rangle\langle\bm{n}_{k}\rvert
U_{kj}\rvert\bm{n}_{j}\rangle\right].
\end{equation}
In the continuum limit, the unitary matrices
$U_{kj}$ become infinitesimal unitary
transformations generated by an SU(2) matrix
$A(t)$:
\begin{eqnarray}
j & \rightarrow & \bm{x}(\tau),\nonumber \\
k & \rightarrow & \bm{x}(t)+\dot{\bm{x}}(\tau) d\tau,\\
U_{kj} & \rightarrow & e^{i A(\tau) d\tau} =  I+i A(\tau) d\tau +
{\cal O}(d\tau^2), \nonumber
\end{eqnarray}
where
\begin{equation}
A(\tau) = \frac{1}{2}  \sigma^a A^a_{\mu}   \dot {x}_{\mu}(\tau)
\end{equation}
is the tangential component of the $SU(2)$ gauge potential.

After noting that,
\begin{widetext}
\begin{equation}
\langle\bm{n}_{{\bm x}(\tau + d\tau)} \rvert  e^{i A(\tau) d\tau} \rvert \bm{n}_{{\bm x}(\tau)} \rangle  =
 1+  [ i \langle \bm{n}_{{\bm x}(\tau)}   \rvert A(t) \rvert  \bm{n}_{{\bm x}(\tau)}   \rangle
\nonumber \\
- \langle \bm{n}_{{\bm x}(\tau)}   \rvert  \partial_{\tau} \rvert \bm{n}_{{\bm x}(\tau)}  \rangle ] d\tau +
{\cal O}(d\tau^2),
\end{equation}
\end{widetext}
and following the same steps that appear in the derivation of Eq.~\eqref{Berry-for}, we obtain
\begin{eqnarray}
\gamma_{{\cal C}} = i \oint_{{\cal C}} \langle\bm{n}\rvert {\cal
D}_{\mu} \rvert\bm{n}\rangle dx_{\mu} \equiv  \oint_{{\cal C}} {\cal
A}_{\mu} dx^{\mu}. \label{eq:berry}
\end{eqnarray}
Here we have introduced the covariant derivative,
\begin{equation}
{\cal D}_{\mu} \equiv \partial_{\mu} - \frac{i \sigma^a}{2} A^a_{\mu},
\end{equation}
and the covariant Berry connection, 
\begin{equation}
{\cal A}_{\mu} = i \langle\bm{n}\rvert \partial_{\mu}
\rvert\bm{n}\rangle + {1\over 2} n^a A^a_{\mu},
\end{equation}
to make the gauge invariance of $\gamma_{{\cal C}}$ more explicit.

We can now use  Stokes theorem to convert the closed integral of Eq.~\eqref{eq:berry} into an integral over the area $S_{\cal C}$:
\begin{eqnarray}
\oint_{{\cal C}} {\cal A}_{\mu} dx^{\mu} &=& \int_{S_{{\cal C}}} [
\partial_{\mu} {\cal A}_{\nu} - \partial_{\nu} {\cal A}_{\mu} ]
d^{2}\sigma^{\mu\nu}
\nonumber \\
&=&\frac{1}{2} \int_{S_{{\cal C}}} [\partial_{\mu}({n}^{a}A_{\nu}^{a})-\partial_{\nu}({n}^{a}A_{\mu}^{a})]d^{2}\sigma^{\mu\nu}
\nonumber \\
&+&
 \frac{1}{2} \int_{S_{{\cal C}}}\bm{n}\cdot[\partial_{\mu}\bm{n}\times\partial_{\nu}\bm{n}]d^{2}\sigma^{\mu\nu}.
\label{eq1}
\end{eqnarray}
The second term is the contribution \eqref{eq:berry_curvature} that
we derived for the SU(2) invariant case. While this term is zero for
collinear or coplanar magnetic orderings, the first term can still
be finite, implying that collinear and magnetic textures can
generate an effective magnetic field if the SOC is finite. In these
cases, the effective magnetic field is $ B_{\eta} =
\epsilon_{\eta \mu \nu} \partial_{\mu} (n^a A^a_{\nu})/2$. In other
words, the effective U(1) vector potential is obtained by projecting
the SU(2) vector potential into the direction of the ${\bm n}$
field.

Our next goal is to find a covariant form for the two contributions that appear in Eq.~\eqref{eq1}.
The integral that appears in the first term can be reexpressed as
\begin{eqnarray}
&&\int_{S_{{\cal C}}}
[\partial_{\mu}(n^{a}A_{\nu}^{a})-\partial_{\nu}(n^{a}A_{\mu}^{a})]d^{2}\sigma^{\mu\nu}
\nonumber \\
 & = & \int_{S_{{\cal C}}}[n^{a}\partial_{\mu}A_{\nu}^{a}-n^{a}\partial_{\nu}A_{\mu}^{a}+(\partial_{\mu}n^{a})A_{\nu}^{a}-(\partial_{\nu}n^{a})A_{\mu}^{a}]d^{2}\sigma^{\mu\nu}
 \nonumber \\
 & = & \int_{S_{{\cal C}}}[n^{a}F_{\mu\nu}^{a}+n^{a}\varepsilon^{abc}A_{\mu}^{b}A_{\nu}^{c}+ A_{\nu}^{a}\partial_{\mu}n^{a} - A_{\mu}^{a} \partial_{\nu}n^{a} ]d^{2}\sigma^{\mu\nu},
 \nonumber \\
 \label{eq2}
\end{eqnarray}
where
\begin{eqnarray}
F_{\mu\nu}^{a} & = &
\partial_{\mu}A_{\nu}^{a}-\partial_{\nu}A_{\mu}^{a}-\varepsilon^{abc}A_{\mu}^{b}A_{\nu}^{c}
\label{fmunu}
\end{eqnarray}
refers to the non-Abelian field strength.

When applied to the vector field ${\bm n}$, the
covariant derivative takes the form
\begin{equation}
D_{\mu} = \partial_{\mu} - L^a A^a_{\mu},
\end{equation}
where ${\bm L}$ is the vector of SO(3) generators introduced in Eq.~\eqref{eq-soc-R}. The natural covariant extension of the solid angle density is:
\begin{widetext}
\begin{eqnarray}
 \bm{n}\cdot (D_{\mu}\bm{n}\times D_{\nu}\bm{n} )  &=&  \varepsilon^{abc}n^{a}\left(\delta^{bn}\partial_{\mu}n^{n}-\varepsilon^{bmn}A_{\mu}^{m}n^{n}\right)\left(\delta^{cl}\partial_{\nu}n^{l}-\varepsilon^{ckl}A_{\nu}^{k}n^{l}\right)
 \nonumber \\
&=&  \varepsilon^{abc}n^{a} \partial_{\mu}n^{b}
\partial_{\nu}n^{c} -
\varepsilon^{abc}\varepsilon^{bmn}n^{a}A_{\mu}^{m}n^{n}
\partial_{\nu}n^{c}
 -\varepsilon^{abc}\varepsilon^{ckl}n^{a} \partial_{\mu}n^{b} A_{\nu}^{k}n^{l}+\varepsilon^{abc}\varepsilon^{bmn}\varepsilon^{ckl}n^{a}A_{\mu}^{m}A_{\nu}^{k}n^{n}n^{l}
 \nonumber \\
 &=&  \varepsilon^{abc}n^{a} \partial_{\mu}n^{b}  \partial_{\nu}n^{c} - \partial_{\nu}n^{c} A_{\mu}^{c}+ \partial_{\mu}n^{b}
 A_{\nu}^{b}+\varepsilon^{ckl}A_{\mu}^{c}A_{\nu}^{k}n^{l}.
\end{eqnarray}
\end{widetext}
By combining this equation with Eqs.~\eqref{eq1}, \eqref{eq2}, and
\eqref{fmunu}, we find a concise covariant form of the Berry phase:
\begin{eqnarray}
\gamma_{{\cal C}} &=& \frac{1}{2} \int_{S_{{\cal
C}}}[n^{a}F_{\mu\nu}^{a}+\bm{n}\cdot(D_{\mu}\bm{n}\times
D_{\nu}\bm{n})]d^{2}\sigma^{\mu\nu}. \label{Berry-fin}
\end{eqnarray}
This final equation is one of the key contributions of this work.
This equation simply tells us that the strength of the effective
U(1) gauge field that is generated by the localized magnetic moments
is the sum of the covariant scalar spin chirality and the the
projection of the SU(2) field strength along the
local direction ${\bm n}$ of the localized
moments.~\cite{Gubarev02,Gubarev04} This is essentially the same
result that was obtained in Eq.~\eqref{llwll} by working on the
lattice and taking the long wavelength limit at the end of the
process.

Another interesting aspect of this derivation is that it can be
immediately generalized to the case of  time-dependent magnetic
configurations by allowing the Greek indices $\mu$ and $\nu$ to run
from $0$ to $d$, where $0$ is the time coordinate and the $d$ is the
spatial dimension of the system under consideration. The zeroth
component of the SU(2) vector potential arises from a Zeeman
coupling between the spin of the conduction electrons and an
external magnetic field ${\bm H}$,
\begin{equation}
A^a_0=   - \frac{g \mu_B}{2} H^a,
\end{equation}
where $g$ is the $g$-factor of the electron and $\mu_B$ is the Bohr
magneton. The action that results from adding the Zeeman
term to the effective Hamiltonian
in Eq.~\eqref{Hamiltonian},
\begin{equation}
S[\Psi] = \int dt (d{\bm x}  \Psi^{\dagger} i \partial_t \Psi -
{\cal H}),
\end{equation}
preserves the invariance under time dependent transformations of the local spin reference frame,~\cite{Tokatly08}
\begin{equation}
\Psi \to {\cal U} \Psi, \;\;\; A^a_{\mu} \sigma^a \to {\cal U} A^a_{\mu} \sigma^a {\cal U}^{-1} -
2 i (\partial_{\mu} {\cal U}) {\cal U}^{-1},
\label{gtransf}
\end{equation}
where $\Psi$ is the electronic wave function in the continuum and
${\cal U}= e^{i \theta^a({\bm x}, t) \sigma^a/2}$.
Eq.~\eqref{gtransf} generalizes the time-independent rotations of
the local reference frame in
Eqs.~\eqref{lsre}-\eqref{guu} that were introduced for the lattice
Hamiltonian.  The {\it effective} electromagnetic field tensor
produced by a time-dependent configuration of the local magnetic
moments is then given by
\begin{equation}
\frac{1}{2} [n^{a}F_{\mu\nu}^{a}+\bm{n}\cdot(D_{\mu}\bm{n}\times
D_{\nu}\bm{n})],
\end{equation}
where $0 \leq \mu, \nu \leq d$ and the strength of the SU(2) gauge
field is still given by Eq.~\eqref{fmunu}. This
equation generalizes then well-known result for SU(2) invariant
systems.~\cite{Volovik03}

\section{Momentum space Berry curvature} \label{sec-mom}

In this section we will discuss a few examples to
apply the notion of the generalized Berry
curvature that was introduced in previous sections. In particular,
we will consider an extended version of the model Hamiltonian
\eqref{Hamiltonian} that includes potentially
anisotropic exchange interactions between the local moments:
\begin{equation}
{\cal H}={\cal H}_{t}+{\cal H}_{J}+{\cal H}_{H}, \label{H2d}
\end{equation}
where ${\cal H}_{t}$ and ${\cal H}_{K}$ have been introduced in Eq.~\eqref{Hamiltonian} and
\begin{eqnarray}
{\cal H}_{H} & = & \sum_{\langle jk  \rangle}  S_{j}^{\mu}{\cal J}^{\mu\nu}_{jk}   S_{k}^{\nu}.
\label{Hamiii}
\end{eqnarray}
We will consider different 2D and 3D variants of this model
comprising  single  or vertically stacked Kagome
layers. The first case corresponds to a very simple version of the
model, which is useful for illustrating the connection between real
space Berry curvature introduced in the previous sections and the
resulting momentum space Berry curvature and AHE.

\begin{figure}[!htb]
\includegraphics[width=7cm]{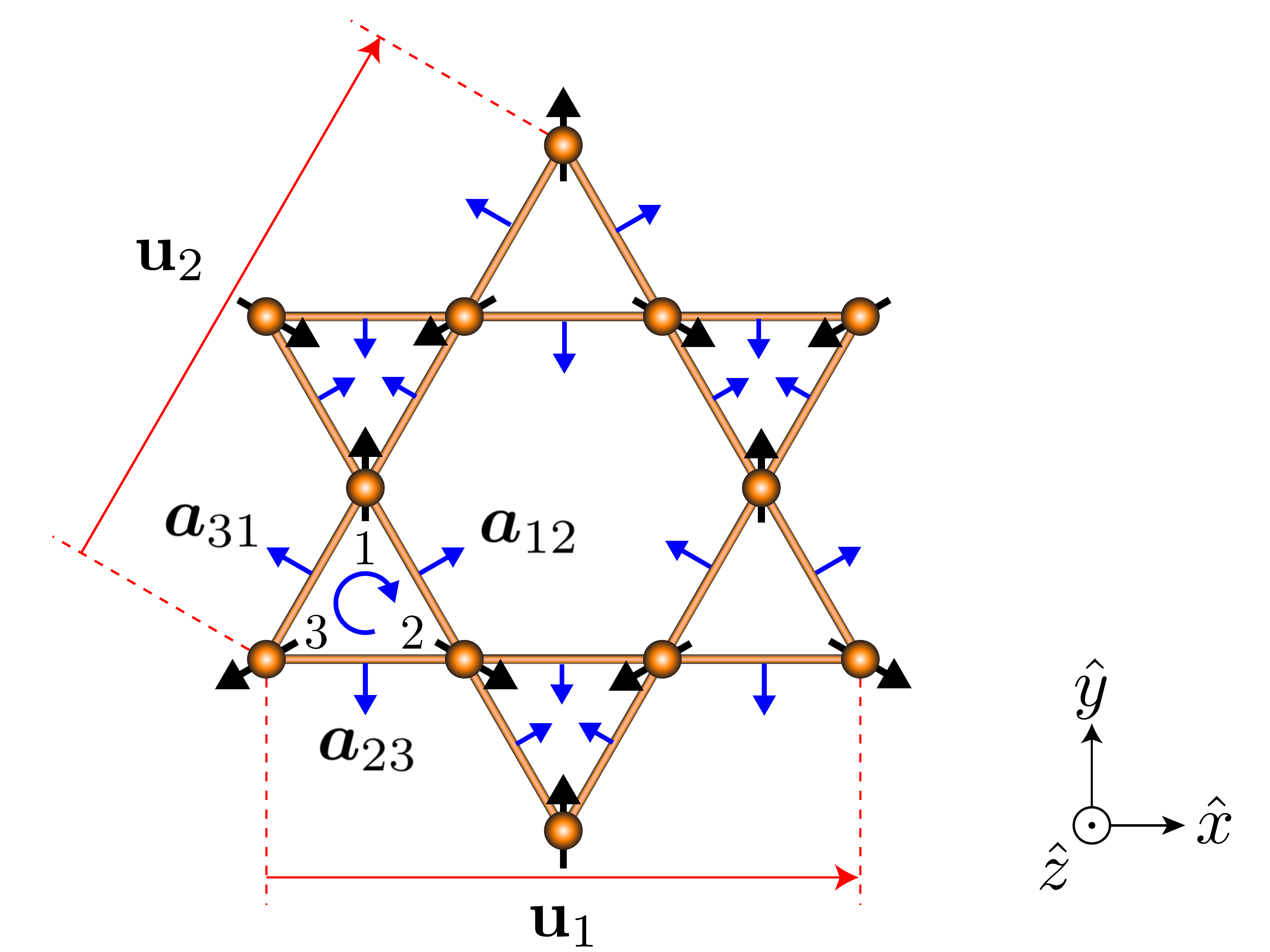}
\caption{Uniform $120^{\circ}$ magnetic ordering
(black arrows) on the Kagome lattice. This is a straightforward
extension of the single triangle state  shown in
Figure~\ref{triang}. The clockwise circulation indicates the bond
orientation. The in-plane components of the SOC vectors ${\bm
a}_{12}^{\bot}$, ${\bm a}_{23}^{\bot}$, ${\bm a}_{31}^{\bot}$ are
indicated with blue arrows. ${\bm u}_{1,2}$ are the primitive
lattice vectors of the Bravais lattice.} \label{Kagome}
\end{figure}

\subsection{Single Kagome Layer}

We will first assume that ${\cal H}_{t}$ is a tight-binding Hamiltonian on a single Kagome layer with nearest-neighbor hopping $t$  and that ${\cal H}_{H} $ stabilizes the ground state magnetic  ordering shown in Fig.~\ref{Kagome}: ${\bm{S}}_{j}=S{\bm{n}}_{j}$, with
\begin{equation}
{\bm{n}}_{1}  =  {\hat{y}},
\;\;\;
{\bm{n}}_{2}  =  \frac{\sqrt{3}{\hat{x}}}{2}-\frac{{\hat{y}}}{2},
\;\;\;
{\bm{n}}_{3} =  \frac{-\sqrt{3}{\hat{x}}}{2}-\frac{{\hat{y}}}{2}.
\end{equation}
The SOC vectors introduced in  Eq.~(\ref{eq-gen-U}) are
\begin{eqnarray}
{\bm{a}}_{23} & = & \cos{\theta}{\bm{a}}_{23}^{\bot}+\sin{\theta}{\hat{z}},\nonumber \\
{\bm{a}}_{31} & = & \cos{\theta}{\bm{a}}_{31}^{\bot}+\sin{\theta}{\hat{z}},\nonumber \\
{\bm{a}}_{12} & = & \cos{\theta}{\bm{a}}_{12}^{\bot}+\sin{\theta}{\hat{z}},
\end{eqnarray}
where ${\bm a}_{ij}^{\bot}=-\epsilon_{ijk}{\bm n}_{k}$.
 Before proceeding with the actual calculation, it is instructive to analyze the distribution of the real-space Berry curvature in the double-exchange limit $J/t \to \infty$.
For $\theta={\pi \over 2}$, the SOC vectors ${\bm a}_{ij}$ are
collinear and  parallel to the ${\hat z}$-axis. The Berry phase
$\gamma_{123}$ picked up by an electron that moves around the
triangle $123$ (see Fig.~\ref{Kagome}) is equal to $\pi$ (the same
is true for the hexagonal plaquettes). ~\footnote{From the geometric
approach, the Berry phase $\gamma_{123}$ is equal to the solid angle
of the spherical \textit{nonagon} spanned by ${\bm n}_1$, ${\bm
v}_{12}$, ${\bm w}_{12}$, ${\bm n}_2$,${\bm v}_{23}$, ${\bm
w}_{23}$, ${\bm n}_3$, ${\bm v}_{31}$, ${\bm w}_{32}$. In the
present example, the nine unit vectors are coplannar, implying that
the solid angle is $2\pi$ and  $\gamma_{123}=\pi$. The same result
can be obtained using the  algebraic approach [see
Eqs.~(\ref{Berrysol})-(\ref{delomegajkl})].} Given that these
$\pi$-phases remain invariant under time-reversal symmetry, there is
no effective magnetic field in real space. The
tight-binding spectrum of the double-exchange
Hamiltonian exhibits two Dirac points at the K points of the
Brillouin zone (BZ): $(\pm  {4\pi \over 3 u}, 0)$ with $u \equiv
\rvert {\bm u}_1 \rvert$. In the opposite limit of in-plane SOC
vectors ($\theta=0$), the Berry phase becomes $\gamma_{123} = \pi -
6 \alpha$ to  first order in the strength of the SOC
($\alpha_{kj}\equiv \alpha$). The resulting effective magnetic flux
per triangle gaps out the  two Dirac points, giving rise  to a
finite Chern number $C = -\text{sign}(\alpha)$ of the lower band of
the massive Dirac fermions.

In the more general case (away from the double-exchange limit), we
need to consider six bands. The matrix of ${\cal
H}_{t}+{\cal H}_{J}$ in momentum space is the $6\times6$ matrix:
\begin{equation}
[H_{\bm{k}}]=\left[\begin{array}{ccc}
H_{\bm{k}}^{11} & H_{\bm{k}}^{12} & (H_{\bm{k}}^{31})^{\dagger}\\
(H_{\bm{k}}^{12})^{\dagger} & H_{\bm{k}}^{22} & H_{\bm{k}}^{23}\\
H_{\bm{k}}^{31} & (H_{\bm{k}}^{23})^{\dagger} & H_{\bm{k}}^{33}
\end{array}\right]\label{eq:tight_binding}
\end{equation}
with
\begin{eqnarray}
H_{\bm{k}}^{12} & \! =\!  &  t e^{i {\alpha \over 2}{\bm{\sigma}}\cdot{\bm{a}}_{12}} (1+e^{-i{\bm{k}}\cdot({\bm{u}}_{2}-{\bm{u}}_{1})}),
\nonumber \\
H_{\bm{k}}^{23} &\! =\! & te^{i {\alpha \over 2} {\bm{\sigma}}\cdot{\bm{a}}_{23}} (1+e^{-i{\bm{k}}\cdot{\bm{u}}_{1}}),
H_{\bm{k}}^{31} \! =\!   te^{i {\alpha \over 2} {\bm{\sigma}}\cdot{\bm{a}}_{31}}(1+e^{i{\bm{k}}\cdot{\bm{u}}_{2}}),
\nonumber \\
H_{\bm{k}}^{11} &\! =\! &- \frac{JS}{2} {\bm{n}}_{1}\! \cdot \!{\bm{\sigma}},
H_{\bm{k}}^{22} \! =\!  - \frac{JS}{2} {\bm{n}}_{2}\!\cdot\!{\bm{\sigma}},  H_{\bm{k}}^{33}\! =\! - \frac{JS}{2} {\bm{n}}_{3}\! \cdot \!{\bm{\sigma}}.
\nonumber
\end{eqnarray}
In the absence of SOC ($\alpha=0$), the tight-binding model hosts
multiple Dirac points. As explained above, these Dirac points are
gapped out for finite SOC ($\alpha\neq0$) as long as the in-plane
component of ${\bm a}_{ij}$ is non-zero. The resulting energy bands
become topologically non-trivial, namely, they acquire a finite
Chern number. We note that this result is consistent with our simple
analysis of the double-exchange limit. As an example, for $J=0.5t$,
$\alpha=0.2 \pi$ and $\theta=0$, the Chern numbers of each band are
$C^{(n=1)}  =  1$, $C^{(n=2)}=-3$, $C^{(n=3)}=2$, $C^{(n=4)}  =  2$,
$C^{(n=5)}=-3$ and $C^{(n=6)}=1$, where $n=1$ ($n=6$) refers to the
lowest (highest) energy band. The resulting energy bands for this
example are shown in Fig.~\ref{spectrum}. Note, however, that these
bands are not adiabatically connected with the six bands that are
obtained in the double-exchange limit (three low-energy bands
separated by an energy $J$ from the three high-energy bands) because
the Chern numbers in the double-exchange limit  must satisfy the
property $C^{(n)}  = - C^{(n+3)}$ for $1 \leq n \leq 3$. This is
just a consequence of the opposite sign of the real space Berry
curvature for bands of opposite spin, i.e., aligned or anti-aligned
with the local moments.

\begin{figure}[!htb]
\includegraphics[width=\columnwidth]{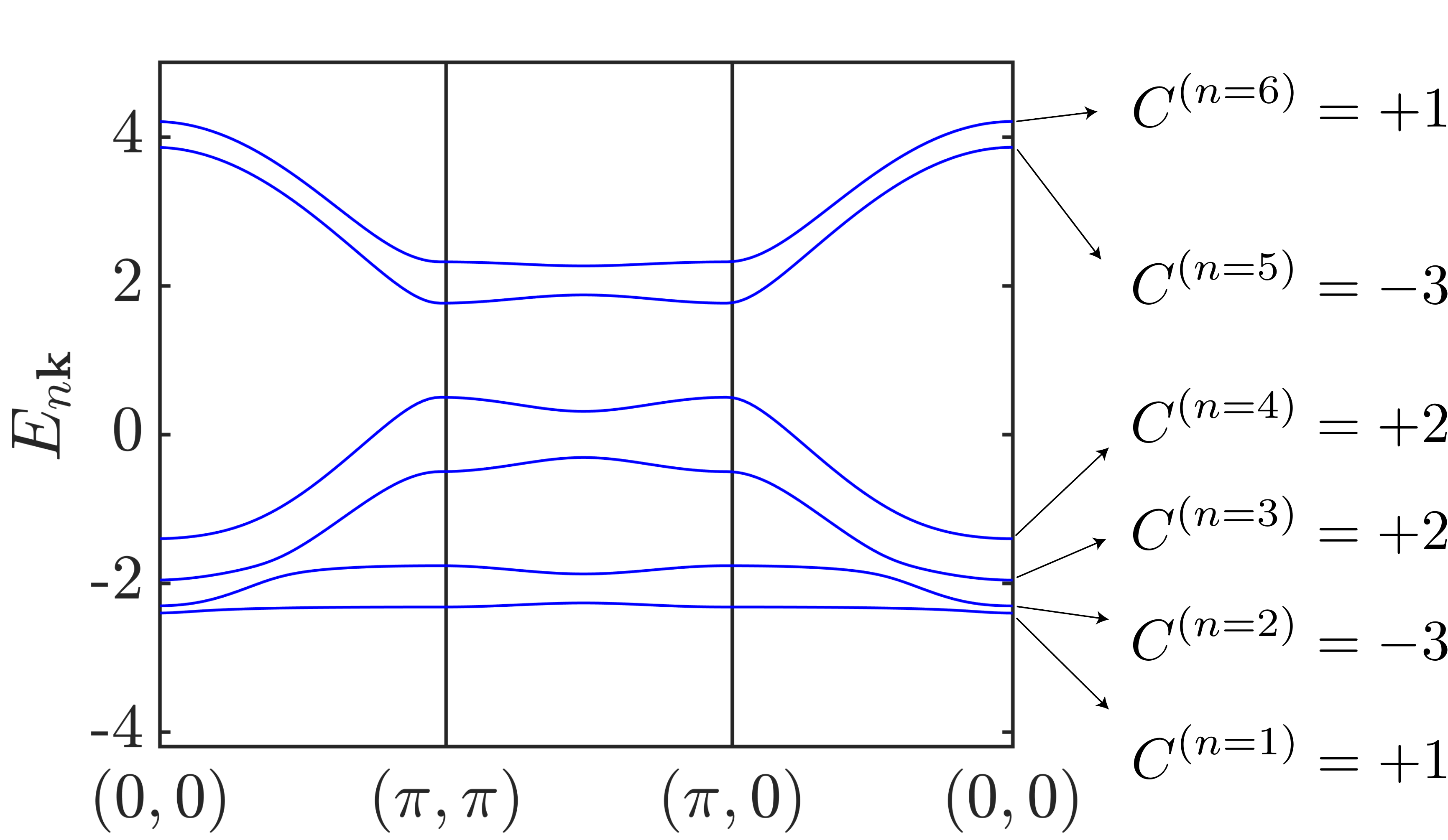}
\caption{Band structure of the tight-binding model of Eq.~(\ref{eq:tight_binding}) for Hamiltonian
parameters $J=0.5t$, $\alpha=0.2\pi$, $\theta=0$.}
\label{spectrum}
\end{figure}

\subsection{Vertically stacked Kagome layers}

We first consider a toy model of vertically stacked Kagome layers
with an SU(2) invariant hopping between adjacent layers. The
intra-layer magnetic ordering is assumed to be the same uniform
$120^{\circ}$ structure that we used in the
previous 2D analysis, while the inter-layer ordering is assumed to
be ferromagnetic. For concreteness, we will assume that the
inter-layer hopping is finite only between nearest-neighbor
($t_{z}^1$) and next-nearest-neighbor ($t_{z}^2$) sites on adjacent
Kagome layers. The resulting inter-layer Hamiltonian in momentum
space is:
\begin{equation}
\tilde{H}^{\rm inter}_{\bm{k}}=\left[\begin{array}{ccc}
\tilde{H}_{\bm{k}}^{11} & \tilde{H}_{\bm{k}}^{12} & (\tilde{H}_{\bm{k}}^{31})^{\dagger}\\
(\tilde{H}_{\bm{k}}^{12})^{\dagger} & \tilde{H}_{\bm{k}}^{22} & \tilde{H}_{\bm{k}}^{23}\\
\tilde{H}_{\bm{k}}^{31} & (\tilde{H}_{\bm{k}}^{23})^{\dagger} & \tilde{H}_{\bm{k}}^{33}
\end{array}\right],\label{eq:tight_binding}
\end{equation}
\noindent where
\begin{eqnarray}
\tilde{H}_{\bm{k}}^{12} & = & 2t_{z}^{2}\cos({\bm{k}}\cdot{\bm{u}}_{3})(1+e^{-i{\bm{k}}\cdot({\bm{u}}_{2}-{\bm{u}}_{1})}),\nonumber \\
\tilde{H}_{\bm{k}}^{23} & = & 2t_{z}^{2}\cos({\bm{k}}\cdot{\bm{u}}_{3})(1+e^{-i{\bm{k}}\cdot{\bm{u}}_{1}}),\nonumber \\
\tilde{H}_{\bm{k}}^{31} & = & 2t_{z}^{2}\cos({\bm{k}}\cdot{\bm{u}}_{3})(1+e^{i{\bm{k}}\cdot{\bm{u}}_{2}}),\nonumber \\
\tilde{H}_{\bm{k}}^{11} & = & 2t_{z}^{1}\cos({\bm{k}}\cdot{\bm{u}}_{3}),\;\tilde{H}_{\bm{k}}^{22}=\tilde{H}_{\bm{k}}^{11},\;\tilde{H}_{\bm{k}}^{33}=\tilde{H}_{\bm{k}}^{11},
\nonumber
\end{eqnarray}
and ${\bm{u}}_{3}$ is the primitive lattice basis vector along the $c$-axis.

We will start by considering the trivial 2D limit of zero
inter-layer hopping. In this limit the spectrum does not depend on
$k_z$, i.e.,  it is the same for  each two-dimensional layer
$(k_x,k_y)$ in momentum space. We have seen that, in
the absence of SOC, each $(k_{x},k_{y})$ layer
hosts several Dirac points, which become nodal lines in the 3D BZ.
These nodal lines are fully gapped out by a finite SOC, implying
that we can introduce a $k_z$-independent Chern number on any
$(k_x,k_y)$ plane, $C^{(n)}(k_{z})$, for  each of the six bands.

The band structure obtained in the 2D limit changes qualitatively
for finite inter-layer hopping because the Dirac lines are gapped
everywhere, except for isolated points that turn out to be Weyl
points of the 3D band structure. The finite inter-layer hopping
leads to a $k_{z}$-dependence of the single-particle dispersion and,
consequently, of the Chern number $C^{(n)}(k_{z})$. The Chern number
remains well-defined unless the $(k_x,k_y)$ layer crosses the Weyl
points and the spectrum is thus gapless. In terms
of momentum space Berry curvature, the Weyl points are magnetic
monopoles, i.e., sources and sinks of the momentum space Berry
curvature. From Gauss's law, we get that the difference between the
$n^{\rm th}$ band Chern numbers at the $k_z$ and $k_z^{\prime}$
planes  is equal to the sum of the charges of the
Weyl points connected to that band which are enclosed by the two
planes . Fig.~\ref{Chern_3D} shows an example of $C^{(n)}(k_{z})$,
where each Kagome layer is identical to that in Fig.~\ref{spectrum},
and a finite inter-layer hopping is included. Only the  Chern
numbers of the lower and the upper two bands ($n=1,2,5,6$) are
changing because the $n=2$ and $n=3$ bands are not connected to any
Weyl point.

\begin{figure}[!htb]
\hspace*{-0.7cm}
\includegraphics[width=10cm]{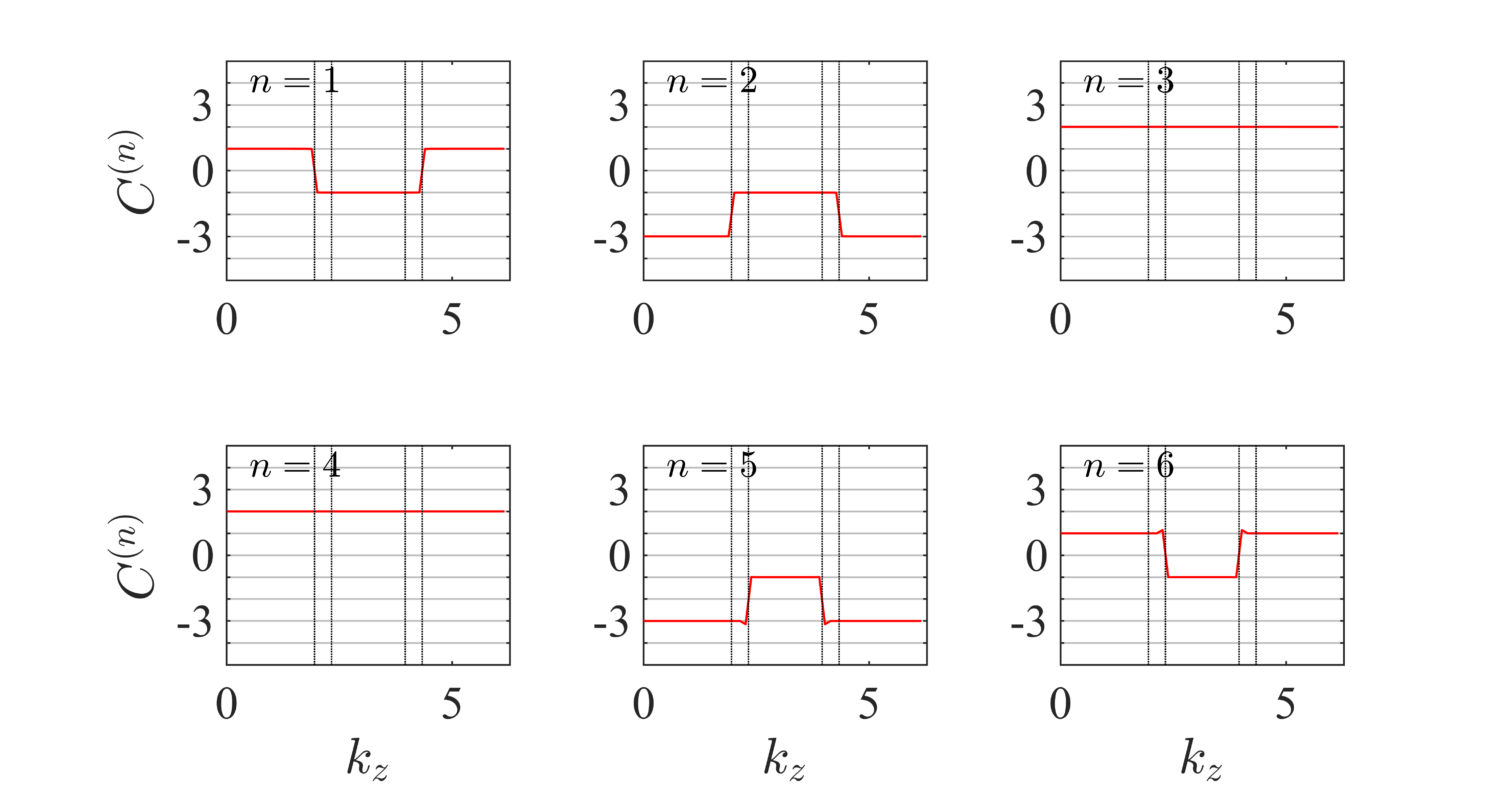}
\caption{Chern number $C(k_z)$ for each $(k_x,k_y)$
plane in momentum space. The in-plane parameters
are the same as in Fig.~\ref{spectrum}, while the  inter-layer
hopping amplitudes are $t_{z}^{1}=0.1t$ and $t_{z}^{2}=0.3t$.}
\label{Chern_3D}
\end{figure}

\subsection{Minimal model for Mn$_{3}$Sn}

We will consider now a modified version of the previous model that
can be regarded as  a minimal Hamiltonian for the lattice and
magnetic ordering of
Mn$_{3}$Sn.~\cite{nakatsuji2015large,ito2017anomalous}   While this
Hamiltonian is not a realistic model for Mn$_{3}$Sn, it includes the
essential ingredients that are required for capturing the
qualitative behavior of this material. More specifically, the model
reveals the origin of the real space Berry curvature that leads to
the  Weyl points that are obtained by more realistic band structure
calculations~\cite{kubler2014non,ito2017anomalous,kubler2018weyl}
and that are the sources and drains of momentum space Berry
curvature.  The crystallographic unit cell of Mn$_{3}$Sn is shown in
Fig.~\ref{fig:mn3sn}~(a): each unit cell includes
six Mn atoms distributed in two Kagome layers. As shown in Fig.~\ref{fig:mn3sn}~(b), the $P6_{3}/mmc$
space group of this material  includes 
\begin{itemize}
\item Mirror symmetries ${\cal M}_{i}$, $i=1,2,3$;
\item Glide symmetries ${\cal G}_i \equiv {\cal M}_{i}^{\prime} \otimes {\cal T}_{1\over 2}$, $i=1,2,3$,
with ${\cal T}_{1\over 2}$ the translation along $c$-axis by $\frac{1}{2}\bm{u}_{3}$;
\item Mirror symmetry ${\cal M}_{z}$ about the Kagome layer;
\item Inversion symmetry ${\cal I}$.
\end{itemize}
The experimental data~\cite{nakatsuji2015large} shows that
Mn$_{3}$Sn displays the magnetic ordering depicted in
Fig.~\ref{fig:model}, which is stable within the temperature range
$50$K $< T <$ $T_{N}$, with a N{\' e}el temperature $T_{N}
\simeq420$K. To a good approximation, this  magnetic ordering
consists of a $120^{\circ}$ structure with fixed
vector spin chirality: the spin  rotates anticlockwise when
circulating clockwise around each triangular plaquette. Several
discrete symmetries are spontaneously broken by this magnetic
ordering. The residual symmetry group includes ${\cal M}_{1}$,
${\cal G}_2 \otimes \Theta$, ${\cal M}_{z} \otimes \Theta$, 
and ${\cal I}$, where  $\Theta$ is the time
reversal operation. The $C_{3}$ symmetry is spontaneously broken
because the spin and the lattice must be rotated in {\it opposite
directions} to keep the system invariant.

\begin{figure}
\centering
\includegraphics[width=\columnwidth]{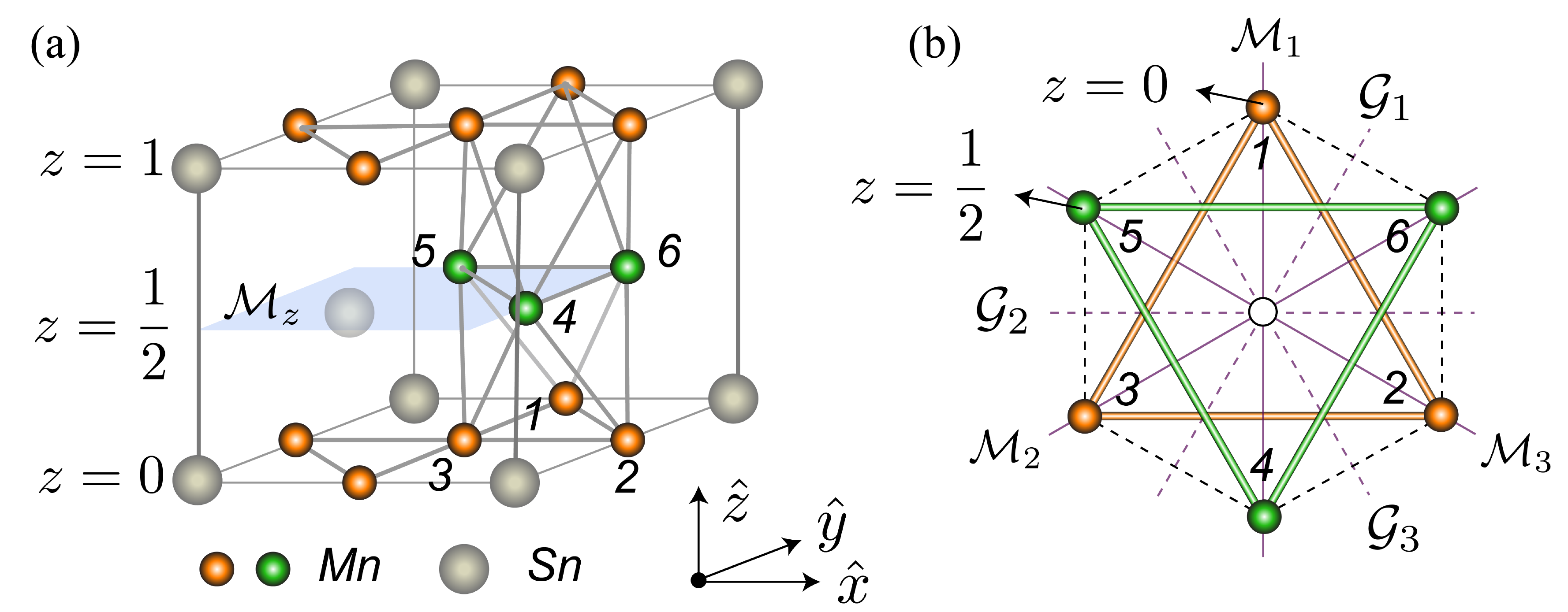}
\caption{(a) Crystallographic unit cell of Mn$_{3}$Sn containing six
Mn atoms   (3 in the $z=0$ plane and 3 in the $1/2$ plane). ${\cal
M}_z$ with $z=1/2$ is a mirror symmetry plane. (b) Vertical mirror planes ${\cal M}_{i}$ and glide planes ${\cal G}_{i}$ ($i=1,2,3$). The white circle at $z=1/4$ indicates an inversion center.}
\label{fig:mn3sn}
\end{figure}


Once again, we will use the minimal Hamiltonian \eqref{Hamiltonian}
to capture the essential features of Mn$_{3}$Sn. The site index $j$
(or $k$) in Eq.~\eqref{Hamiltonian} will be decomposed into two
indices  $(\alpha,\bm{r})$, where, as shown in
Fig.~\ref{fig:mn3sn}~(a), $\alpha=1,...,6$ is a sublattice index and
$\bm{r}$ is the coordinate of the crystallographic unit cell. We
will also assume that the above-mentioned magnetic ordering of the
Mn moments is stabilized by the additional exchange interaction term
that is included in Eq.~\eqref{H2d}. The corresponding orientations
of the magnetic moments on each sublattice are:
\begin{eqnarray}
\bm{n}_{1}  &=& \bm{n}_{4} =  \hat{x}, \nonumber \\
\bm{n}_{2}  &=& \bm{n}_{5} =  -\frac{1}{2}\hat{x}+\frac{\sqrt{3}}{2}\hat{y}, \nonumber \\
\bm{n}_{3} &=& \bm{n}_{6} =  -\frac{1}{2}\hat{x}-\frac{\sqrt{3}}{2}\hat{y}.
\end{eqnarray}

\begin{figure}
\centering
\includegraphics[width=\columnwidth]{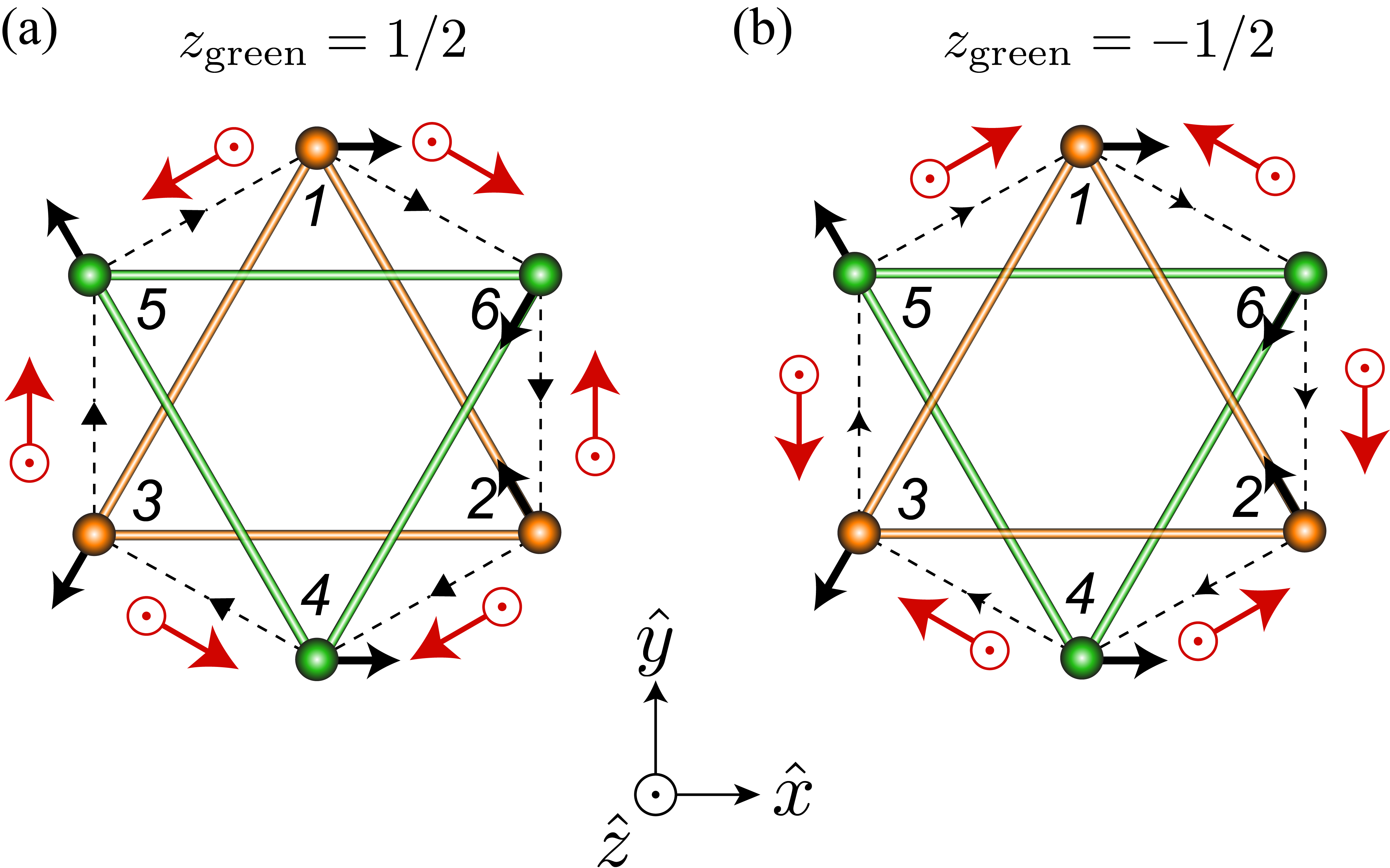}
\caption{Magnetic order (thick black arrows) and spin-orbital
vectors $\bm{a}_{kj}$ (red arrows) on inter-layer bonds connecting
the orange Kagome layer at $z=0$
and the green Kagome layer at (a)
$z={1\over 2}$ and (b) $z=-{1\over 2}$.} \label{fig:model}
\end{figure}

The unit vectors ${\bm a}_{jk}$ are constrained by symmetry considerations.  For instance,
the in-plane components of $\bm{a}_{kj}$ vanish for  bonds lying in each Kagome layer
because of  the mirror symmetry plane ${\cal M}_{z}$, implying that $\bm{a}_{kj} = \alpha \bm{\hat \bm{z}}$.
In addition, as shown in Fig.~\ref{fig:mn3sn}~(a), the two Kagome layers in the unit cell are related by
the inversion symmetry, ${\cal I}$: $j\rightarrow\tilde{j}$,
$k \rightarrow\tilde{k}$, implying that $\bm{a}_{kj}=\bm{a}_{\tilde{k}\tilde{j}}$.
By combining these results with the $C_3$ symmetry  of the lattice, we  obtain that  the six in-plane
hopping matrices must be the same:
\begin{equation}
t_{12}=t_{23}=t_{31}=t_{45}=t_{56}=t_{64} \equiv t  e^{i {\alpha \over 2} \sigma_z}.
\end{equation}

A similar symmetry analysis leads to the  following parametrization
of the inter-layer  hopping matrices:
\begin{eqnarray}
t_{\alpha\beta}^{\prime} & = & t^{\prime}e^{i{\alpha^{\prime}\over 2}\bm{a}_{\alpha\beta}^{\prime}\cdot\bm{\sigma}},\\
t_{\alpha\beta}^{\prime\prime} & = &
t^{\prime}e^{i{\alpha^{\prime}\over
2}\bm{a}_{\alpha\beta}^{\prime\prime}\cdot\bm{\sigma}},
\end{eqnarray}
where $\alpha,\beta$ are sublattice indices connected by inter-layer bonds (see Fig.~\ref{fig:model}) and 
$t_{\alpha\beta}^{\prime}$ ($t_{\alpha\beta}^{\prime\prime}$) are the hopping amplitudes on the inter-layer bonds depicted in Fig.~\ref{fig:model}~(a) (Fig.~\ref{fig:model}~(b)).  The corresponding spin-orbit vectors,
\begin{eqnarray}
\bm{a}_{\alpha\beta}^{\prime} & = & \pm\cos{\theta^{\prime}} {\bm{a}}_{\alpha\beta}^{\bot}+\sin{\theta^{\prime}}\hat{\bm{z}},\\
\bm{a}_{\alpha\beta}^{\prime\prime} & = & \mp\cos{\theta^{\prime}}{\bm{a}}_{\alpha\beta}^{\bot}+\sin{\theta^{\prime}} \hat{\bm{z}},
\end{eqnarray}
are constrained by the mirror symmetries ${\cal M}_{1,2,3}$ and ${\cal M}_{z}$.
The upper (lower)  sign corresponds to $\alpha=1,2,3$
($\alpha=4,5,6$), and the vector ${\bm{a}}_{\alpha\beta}^{\bot}$ refers
to the  (normalized) projection of the bond vector
${\bm{a}}_{\alpha\beta}$ on the basal plane
\begin{align}
{\bm{a}}_{16}^{\bot} & =(\frac{\sqrt{3}}{2},-\frac{1}{2},0),\;{\bm{a}}_{24}^{\bot}=(-\frac{\sqrt{3}}{2},-\frac{1}{2},0),\;{\bm{a}}_{35}^{\bot}=(0,1,0),\nonumber \\
{\bm{a}}_{62}^{\bot} & ={\bm{a}}_{35}^{\bot},
\;\;\;\;\;\;\;\;\;\;\;\;\;\;
{\bm{a}}_{43}^{\bot}={\bm{a}}_{16}^{\bot},
\;\;\;\;\;\;\;\;\;\;\;\;\;\;
{\bm{a}}_{51}^{\bot}={\bm{a}}_{24}^{\bot}.
\end{align}

In momentum space, the electron kinetic term is
${\cal H}_t = \sum_{\bm k} \psi_{\bm k}^{\dagger} H_{\bm k}
\psi_{\bm k} $ with $\psi_{\bm k} = (c_{1{\bm
k}},c_{2{\bm k}},c_{3{\bm k}},c_{4{\bm k}},c_{5{\bm k}},c_{6{\bm
k}})^T$ and
\begin{eqnarray}
H_{\bm k}  = \left[ \begin{array}{cccccc}
H^{11}_{\bm k} & H^{21\dagger}_{\bm k} & H^{13}_{\bm k} & 0 & \tilde{H}^{15}_{\bm k} & \tilde{H}^{61\dagger}_{\bm k}  \\
H^{21}_{\bm k} & H^{22}_{\bm k} & H^{32\dagger}_{\bm k} & \tilde{H}^{24\dagger}_{\bm k} & 0 & \tilde{H}^{26}_{\bm k} \\
H^{13\dagger}_{\bm k} & H^{32}_{\bm k} & H^{33}_{\bm k} & \tilde{H}^{34}_{\bm k} & \tilde{H}^{53\dagger}_{\bm k} & 0  \\
0 & \tilde{H}^{42}_{\bm k} & \tilde{H}^{34\dagger}_{\bm k} & H^{11}_{-\bm k} & H^{21\dagger}_{-\bm k} & H^{13}_{-\bm k} \\
\tilde{H}^{15\dagger}_{\bm k} & 0 & \tilde{H}^{53}_{\bm k} & H^{21}_{-\bm k} & H^{22}_{-\bm k} & H^{32\dagger}_{-\bm k} \\
 \tilde{H}^{16}_{\bm k} & \tilde{H}^{26\dagger}_{\bm k} & 0 & H^{13\dagger}_{-\bm k} & H^{32}_{-\bm k} & H^{33}_{-\bm k} \\
\end{array}\right],
\end{eqnarray}
where the diagonal elements are $H_{\bm k}^{\alpha \alpha} = -{1\over 2} J S {\bm e}_{\alpha}\cdot {\bm \sigma}$, the off-diagonal elements $H^{\alpha \neq \beta}_{\bm k}$ arise from the intra-layer hopping terms,
\begin{eqnarray}
H_{\bm{k}}^{21} & = & t \left(1+e^{-i\bm{k}\cdot(\bm{u}_{2}-\bm{u}_{1})}\right) e^{i {\alpha \over 2} \bm{a}\cdot \bm \sigma},\nonumber \\
H_{\bm{k}}^{32} & = & t \left(1+e^{-i\bm{k}\cdot\bm{u}_{1}}\right) e^{i{\alpha \over 2} \bm{a}\cdot \bm \sigma},\\
H_{\bm{k}}^{13} & = & t \left(1+e^{i\bm{k}\cdot\bm{u}_{2}}\right) t e^{i{\alpha \over 2} \bm{a}\cdot \bm \sigma},\nonumber
\end{eqnarray}
and the off-diagonal elements $\tilde{H}^{\alpha \neq \beta}_{\bm k}$ arise from the inter-layer hopping terms
\begin{eqnarray}
\tilde{H}_{\bm{k}}^{61} & = &
t_{16}^{\prime}+t_{16}^{\prime\prime}e^{i\bm{k}\cdot\bm{u}_{3}},
\,\,
\tilde{H}_{\bm{k}}^{42} =  t_{24}^{\prime}+t_{24}^{\prime\prime}e^{i\bm{k}\cdot\bm{u}_{3}},\nonumber \\
\tilde{H}_{\bm{k}}^{53} & = &
t_{35}^{\prime}+t_{35}^{\prime\prime}e^{i\bm{k}\cdot\bm{u}_{3}},
\,\,
\tilde{H}_{\bm{k}}^{34}  =  t_{43}^{\prime}+t_{43}^{\prime\prime}e^{-i\bm{k}\cdot\bm{u}_{3}},\nonumber \\
\tilde{H}_{\bm{k}}^{15} & = &
t_{51}^{\prime}+t_{51}^{\prime\prime}e^{-i\bm{k}\cdot\bm{u}_{3}},
\,\, \tilde{H}_{\bm{k}}^{26}  =
t_{62}^{\prime}+t_{62}^{\prime\prime}e^{-i\bm{k}\cdot\bm{u}_{3}}.\nonumber
\end{eqnarray}

\begin{figure}
\centering
\includegraphics[width=\columnwidth]{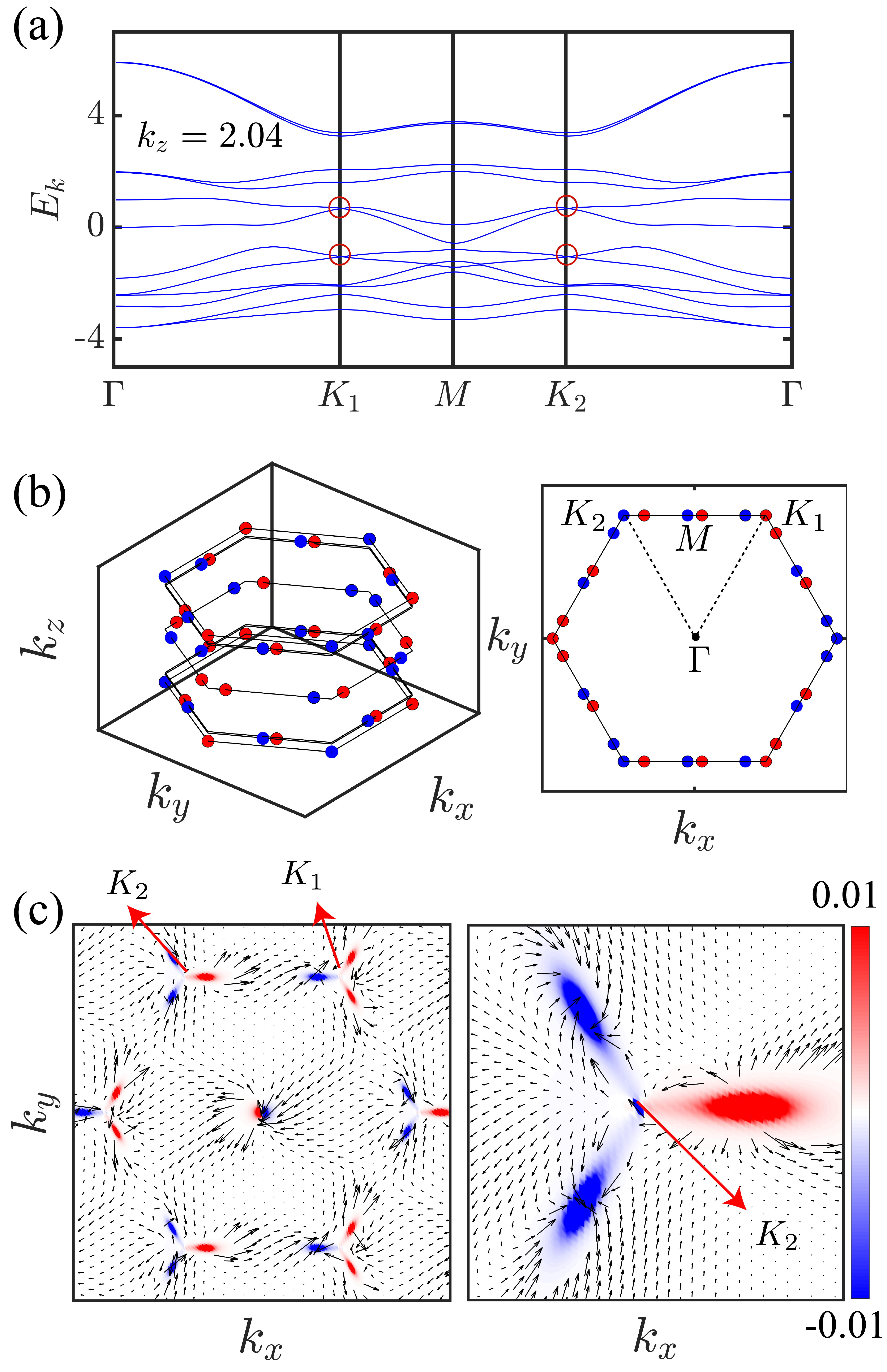}
\caption{(a): Band structure along symmetric paths of the BZ. Red
circles indicate Weyl points. (b) Weyl points in the first BZ. Red
circles correspond to $Q>0$, while blue circles correspond to $Q<0$.
(c) Left: vector field ${\cal {\bm B}}^{(5)}$ associated with the
fifth band on the $q_{z}=0^{+}$ plane (taken as $q_{z}=10^{-4}$ in
numerical calculation). Right: enlarged plot around the $K_{2}$
point. The model parameters are: $t^{\prime}=t$,
$J=2t$, $\alpha=0.2\pi$, $\alpha^{\prime}=0.2\pi$,
and $\theta^{\prime} = \pi/4 \approx 0.785$.}
\label{fig:Weylpoints}
\end{figure}

\subsection{Weyl points}

For concreteness, we will consider the following set of Hamiltonian
parameters: $t^{\prime}=t$, $J=2t$,
$\alpha=0.2\pi$, $\alpha^{\prime}=0.2\pi$, and
$\theta^{\prime}= \pi / 4 \approx 0.785$. While
this set does not correspond to the particular case of Mn$_3$Sn, it
is enough to illustrate the physical origin of the Weyl points that
appear in the band structure of this material.~\cite{Yang17} As
shown in Fig.~\ref{fig:Weylpoints}~(a), there are two Weyl points
located at each of the $K$ points of the BZ, connecting the fifth
and the sixth bands and the seventh and the eight bands  at
$k_{z}=2.04$. In general, the Weyl points
connecting the fifth and the sixth bands turn out to be distributed
over the surface of the BZ,  as shown in
Figs.~\ref{fig:Weylpoints}~(b) and (c). This is not true for the
other bands. Except for equivalent points related by reciprocal
lattice vectors, they are classified according to the symmetries
${\cal M}_{1}$, $ {\cal G}_{2} \otimes \Theta$, ${\cal
M}_{z} \otimes \Theta$, and ${\cal I}$.

Weyl points connecting bands $n$ and $n+1$ are singularities of the vector field
$B_{\rho}^{(n)}=\sum_{\mu\nu}\epsilon_{\rho\mu\nu}{\cal B}_{\mu\nu}^{(n)}$,
where  ${\cal B}_{\mu\nu}^{(n)}$ is
the momentum space Berry curvature of the $n$-band  given by the Kubo formula,~\footnote{(One could also consider the
band $n+1$ instead of $n$, after noticing that ${\cal B}_{\mu\nu}^{(n+1)} = - {\cal B}_{\mu\nu}^{(n)}$ near the Weyl point).}
\begin{align}
{\cal B}_{\mu\nu}^{(n)} & =-2\hbar^{2} \! \sum_{m\neq n} \! \text{Im} \! \left[\frac{\langle\psi_{n\bm{k}}\rvert\partial_{k_{\mu}}{\cal H}\rvert\psi_{m\bm{k}}\rangle\langle\psi_{m\bm{k}}\rvert\partial_{k_{\nu}}{\cal H}\rvert\psi_{n\bm{k}}\rangle}{\left(E_{n\bm{k}}-E_{m\bm{k}}\right)^{2}}\right].
\end{align}
$E_{n\bm{k}}$ is the dispersion relation of the $n^{\rm th}$ band and $\rvert\psi_{n\bm{k}}\rangle$
is the associated Bloch wave function. The singularity at the Weyl point
is characterized by the quantized topological charge of the monopole
\begin{eqnarray}
Q^{(n)} & = & \frac{1}{4\pi}\int_{\Sigma}d^{2}\bm{s}\cdot  \bm{B}^{(n)},
\end{eqnarray}
where $\Sigma$ is a closed surface enclosing the Weyl point. Each
Weyl point is then a  source (sink) of the Berry curvature field
$\bm{B}^{(n)}$ if $Q^{(n)}>0$ ($Q^{(n)}<0$). This charge $Q^{(n)}$
changes sign under mirror and spatial inversion transformations,
while it remains invariant under time-reversal.
Fig.~\ref{fig:Weylpoints} shows the Weyl points with $Q^{(n)}>0$
(red color) and $Q^{(n)}<0$ (blue color). Pairs related by the
residual symmetries ${\cal M}_{1}$, ${\cal
G}_{2} \otimes \Theta$, ${\cal M}_{z}\otimes \Theta$, and
${\cal I}$ have opposite monopole  charges.

The vector field $\bm{B}^{(n=5)}$ on the $k_{z}=0$ plane is shown
in Fig.~\ref{fig:Weylpoints}~(c). Symmetry
restrictions imply that
\begin{itemize}
\item ${\cal M}_{1}:$  $ (k_{x},k_{y},k_{z})\rightarrow  (-k_{x},k_{y},k_{z})$
\begin{equation}
(B_{x}^{(n)},B_{y}^{(n)},B_{z}^{(n)})\rightarrow(B_{x}^{(n)},-B_{y}^{(n)},-B_{z}^{(n)});
\end{equation}
\item ${\cal G}_{2} \otimes \Theta$: $(k_{x},k_{y},k_{z})\rightarrow(-k_{x},k_{y},-k_{z})$
\begin{equation}
(B_{x}^{(n)},B_{y}^{(n)},B_{z}^{(n)})\rightarrow(B_{x}^{(n)},-B_{y}^{(n)},B_{z}^{(n)});
\end{equation}
\item ${\cal M}_{z}\otimes \Theta$: $(k_{x},k_{y},k_{z})\rightarrow(-k_{x},-k_{y},k_{z}),$
\begin{equation}
(B_{x}^{(n)},B_{y}^{(n)},B_{z}^{(n)})\rightarrow(B_{x}^{(n)},B_{y}^{(n)},-B_{z}^{(n)});
\end{equation}
\item ${\cal I}$: ${\bm k}\rightarrow - {\bm k}$, $ {\bm B}^{(n)}\rightarrow {\bm B}^{(n)},$
\end{itemize}
The first two transformations imply that $B_{z}^{(n)}\equiv0$ for
$k_{z}=0$ ($k_{z}$ is fixed at $10^{-4}$ in
Fig.~\ref{fig:Weylpoints}~(c) such that $B_{z}^{(n)}\neq0$ near the
Weyl points). Note that $B_{x}^{(n)}$ remains even under the listed
residual symmetry transformations, while $B_{y}^{(n)},B_{z}^{(n)}$
are odd under some of them (e.g., the mirror symmetry plane ${\cal
M}_{1}$). This observation implies that the Hall conductivities
$\sigma_{zx}$ and $\sigma_{xy}$ must vanish.

At this point it is interesting to ask what is the distribution of
the {\it real space} Berry curvature that leads to the distribution
of {\it momentum space} Berry curvature depicted in
Fig.~\ref{fig:Weylpoints}. This result is shown in
Fig.~\ref{real_space_Berry}. The first observation is that the
effective flux is zero for intra-layer triangles. The flux (Berry
phase) is concentrated on the triangular plaquettes that connect
different layers. The sign of the flux alternates between the two
types of triangles that connect consecutive layers. Identifying the
origin of the real space Berry curvature is potentially useful for
achieving a more efficient control of  response functions, such as
the Hall response, that are strongly influenced by the momentum
space Berry curvature. For instance, Fig.~\ref{real_space_Berry}
indicates that changing the lattice parameter or the SOC of the
\emph{interlayer} triangles is the
correct strategy for controlling momentum space Berry curvature of
Mn$_3$Sn via the modification of its real space Berry curvature.

\begin{figure}
\centering
\includegraphics[width=4.0cm]{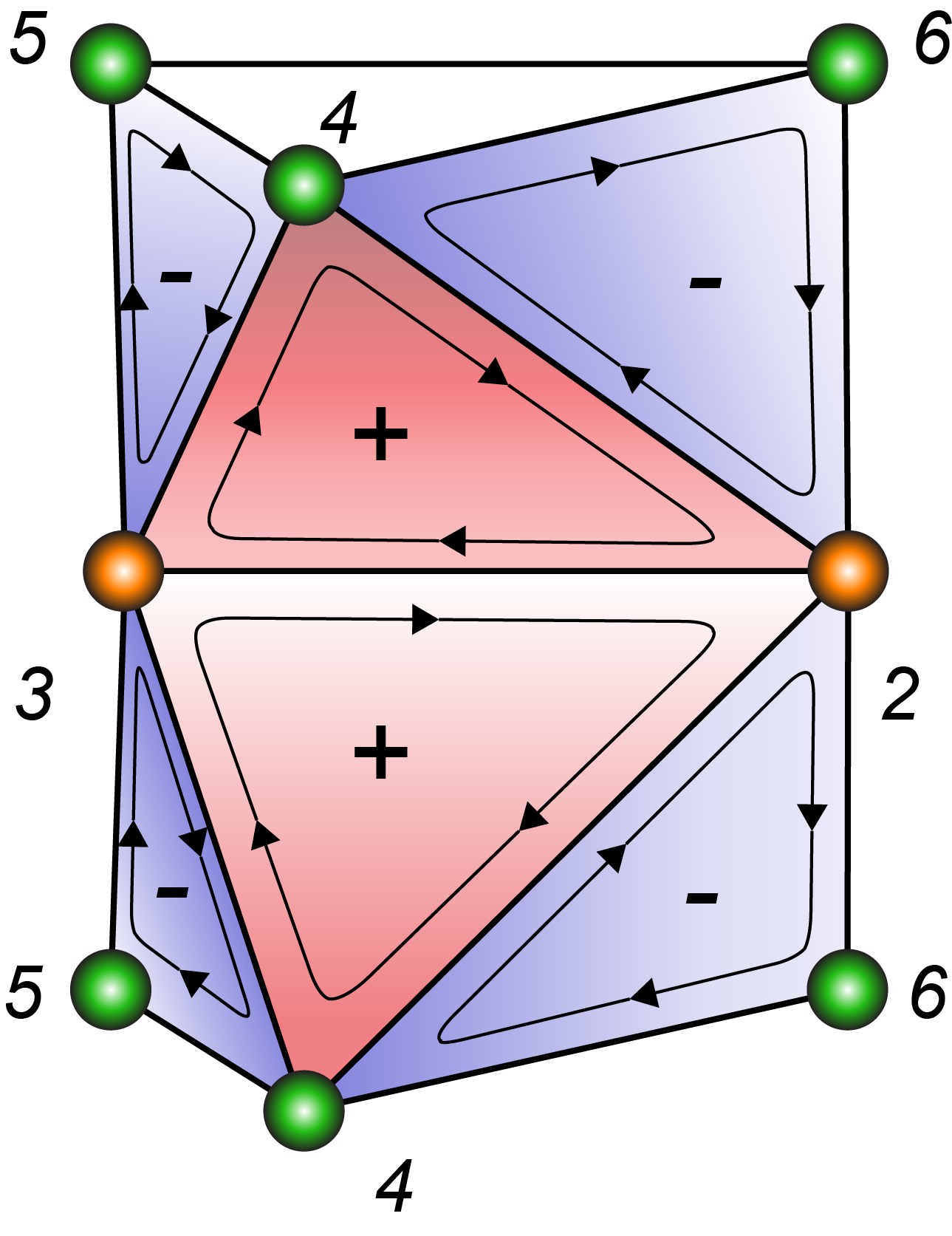}
\caption{ Distribution of real space Berry curvature obtained from the minimal model for Mn$_3$Sn.}
\label{real_space_Berry}
\end{figure}

\subsection{Hall conductivity}

Our final step is to compute the Hall conductivity  produced by the
momentum space Berry curvature. The only purpose of this calculation
is to quantify the order of magnitude of the effect. We note,
however, that a finite Hall conductivity is not the only measurable
consequence of the momentum space Berry curvature. Another potential
consequence  is a finite anomaly-related magnetoresistance  that is
still present in doped Weyl
semimetals.~\cite{Ishizuka19,Ishizuka19b}

The Hall conductivity is given by
\begin{equation}
\sigma_{\mu\nu}=-\frac{e^{2}}{h}\int_{BZ}\frac{d^{3}\bm{k}}{(2\pi)^{3}}\sum_{n} f(E_{n\bm{k}}){\cal B}_{\mu\nu}^{(n)}(\bm{k}),
\end{equation}
where $f(x)=1/(e^{(x-\mu)/k_B T}+1)$ the Fermi-Dirac distribution function. The residual
mirror symmetry plane ${\cal M}_{1}$ leads to $\sigma_{zx}=\sigma_{xy}=0$
The only non-zero component, $\sigma_{yz}$, arises from  the
in-plane component of the spin-orbital vector $\bm{a}_{ij}$.
This is so because the mirror symmetry${\cal M}_{2}$
times  a spin rotation $R_{z}^{(S)}(-\frac{2\pi}{3})$ becomes an element of the residual symmetry group for $\bm{a}_{ij}\parallel\hat{z}$ on every bond.

\begin{figure}
\vspace{0.5cm}
\centering
\includegraphics[width=\columnwidth]{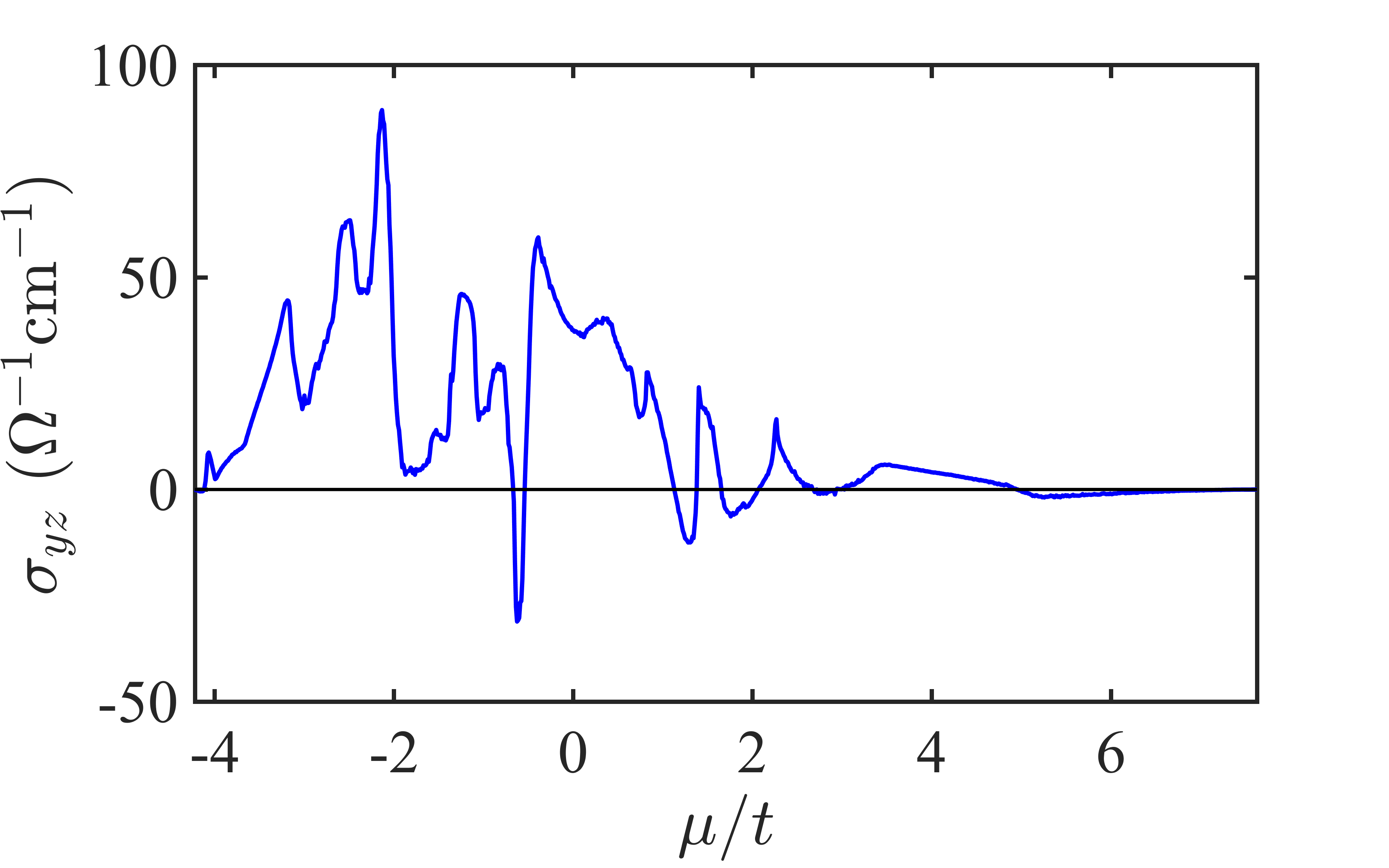}
\caption{Hall conductivity as a function of chemical potential $\mu$ for the same model parameters as in Fig.~\ref{fig:Weylpoints}.}
\label{fig:hall}
\end{figure}

As expected, the value of $\sigma_{yz}$ depends strongly on the
position of the Fermi level $\mu$. However, it is interesting to
note that the order of magnitude of the overall amplitude of the
$\sigma_{yz}(\mu)$ curve coincides with the
experimental value in
Mn$_3$Sn~\cite{nakatsuji2015large}. In other words, our minimal
model for Mn$_3$Sn not only captures the key qualitative aspects of
the problem, but also the correct order of magnitude of response
functions enabled by the finite momentum space Berry curvature. This
attribute of the model can be exploited for further understanding
the interplay between real and momentum space Berry curvature
induced by the combination of magnetic ordering and
SOC.

Finally, we mention that the small uniform magnetization that is
observed in Mn$_3$Sn should have the same origin as the spontaneous
Hall conductivity. The same symmetry analysis that enables a finite
$\sigma_{yz}$ for the antiferromagnetic ordering depicted in
Fig.~\ref{fig:model} also  enables a finite uniform orbital
magnetization along the $x$-axis that should also produce a uniform
spin magnetization along the same direction via the SOC. This effect
can be captured by our minimal model if we allow the
antiferromagnetic  state of Fig.~\ref{fig:model} to relax into the
magnetically ordered state  that minimizes the total energy $\langle
{\cal H}_t + {\cal H}_J + {\cal H}_H\rangle $.~\footnote{Note that
in our previous analysis we assumed for simplicity that the
Heisenberg term is dominant and the optimal magnetic ordering is
then determined by minimization of this term.} As for the case of
Mn$_3$Sn,~\cite{nakatsuji2015large} this uniform magnetization,
which must be present in any antiferromagnet that produces a
spontaneous Hall effect, can be used to orient the antiferromagnetic
domains and induce a net Hall conductivity.

\section{Conclusions}

In summary, we have shown that the real space Berry phase that
electrons pick up when they move in a closed loop while interacting
with local magnetic moments is a geometric property that combines
rotation matrices associated with the finite
SOC and the underlying magnetic ordering. From a more physical
point of view, the finite SOC rotates the electronic spin while the
electron hops from one atomic orbital to another. This rotation
enables a non-trivial Berry phase (different from $0$ or $\pi$)
induced by collinear and coplanar magnetic configurations. In view
of the fact that collinear and coplanar magnetic orderings are more
common than noncoplanar orderings, we can conclude that SOC should
play a crucial role in the discovery of new materials with large
topological Hall effect, or even finite temperature Chern
insulators, induced by spontaneous antiferromagnetic ordering. While
material candidates can be identified by applying a simple symmetry
analysis,~\cite{Chen14,Smejkal19} understanding the underlying
microscopic mechanism for the generation of Berry curvature is of
crucial importance for the optimization and control of the effect.
Understanding the microscopic mechanism is also necessary to
estimate the value of the topological contribution to different
response functions of interest and to anticipate the change of these
response functions under the application of external fields, such as
pressure, strain or magnetic field.

Here we have reduced this microscopic mechanism to its simplest form
by considering the minimal model introduced in
Eq.~\eqref{Hamiltonian}. Like for the SU(2) invariant case of zero
SOC, the mechanism becomes transparent in the double-exchange limit
of this model because the low-energy theory maps into a theory of
spinless fermions  coupled to an effective U(1) gauge field. We have
shown that the strength of this emergent U(1) field
has two covariant contributions. The first contribution  is the
covariant SU(2) extension of the skyrmion density
in the underlying configuration of localized
magnetic moments. The second contribution is simply the projection
of the strength of the SU(2) gauge field produced by the SOC along
the direction of the localized magnetic moments. This simple result
reveals the role of SOC in the generation of real space Berry
curvature. The new scenario becomes particularly clear when the  SOC
can be gauged away by a local rotation of the spin reference frame.
The real space Berry curvature is then equal to the skyrmion density
of the localized magnetic moments in that particular reference
frame. It is then clear that magnetic configurations that are
collinear or coplanar in the laboratory reference frame can become
noncoplanar in the new reference frame.

We have illustrated these concepts by applying them to a simple
minimal model that captures the essential aspects of
Mn$_3$Sn.~\cite{nakatsuji2015large} A similar analysis can in
principle be applied to other materials that exhibit topological
Hall effect induced by coplanar or collinear magnetic
orderings.~\cite{Smejkal19} Moreover, given that the SU(2) gauge
field produced by the SOC is a bond variable (i.e., it depends on
the relative position of the two ions connected by that bond), it is
natural to expect that pressure and strain could play an important
role in the external control of the topological Hall effect. This is
a new control variable enabled by the SOC, in addition to the
external magnetic field that controls the orientation of the
localized magnetic moments via the Zeeman term. It is
then clear that SOC is crucially important for
expanding the spectrum of materials and external
fields that can be used to produce and control potentially large
topological contributions to response functions.

\begin{acknowledgments}
We thank S. Nakatsuji, N. Nagaosa and D. A. Tennant for useful discussions.
Work by H. Zhang was supported by the U.S. Department of Energy (DOE), 
Office of Science, Basic Energy Sciences, Materials Sciences and Engineering Division.
S.-S.~Z.~and C.~D.~B.~are supported by funding from
the Lincoln Chair of Excellence in Physics and from the Los Alamos
National Laboratory Directed Research and Development program.
The work of G.~B.~H.~at ORNL was supported by
Laboratory Director's Research and Development funds.

\end{acknowledgments}

%

\begin{appendix}

\end{appendix}

\end{document}